\documentstyle[12pt,aasms4]{article}

\input epsf.tex
\tightenlines
\slugcomment{Submitted to the Astrophysical Journal}

\def \m{\ifmmode M_\odot\else M$_\odot$\fi} 

\def \lta {\mathrel{\vcenter 
     {\hbox{$<$}\nointerlineskip\hbox{$\sim$}}}} % less than approx
\def \gta {\mathrel{\vcenter 
     {\hbox{$>$}\nointerlineskip\hbox{$\sim$}}}} % greater than approx
\def\vec#1{ {\bf #1} }
\def\cvec#1{ {\cal #1} }

\def\He{\ifmmode {^{4}He} \else $^{4}He$ \fi}
\def\C{\ifmmode {^{12}C} \else $^{12}C$ \fi}
\def\O{\ifmmode {^{16}O} \else $^{16}O$ \fi}
\def\Ne{\ifmmode {^{20}Ne} \else $^{20}Ne$ \fi}
\def\Mg{\ifmmode {^{24}Mg} \else $^{24}Mg$ \fi}
\def\Si{\ifmmode {^{28}Si} \else $^{28}Si$ \fi}
\def\Ar{\ifmmode {^{36}Ar} \else $^{36}Ar$ \fi}
\def\Ca{\ifmmode {^{40}Ca} \else $^{40}Ca$ \fi}
\def\Ti{\ifmmode {^{44}Ti} \else $^{44}Ti$ \fi}
\def\Cr{\ifmmode {^{48}Cr} \else $^{48}Cr$ \fi}
\def\Fe{\ifmmode {^{52}Fe} \else $^{52}Fe$ \fi}
\def\Ni{\ifmmode {^{56}Ni} \else $^{56}Ni$ \fi}

\def\gcc{\ifmmode {\rm g~cm^{-3}} \else g~cm$^{-3}$ \fi}   % density unit

\begin{document}

\title{DEFLAGRATION TO DETONATION TRANSITION IN THERMONUCLEAR SUPERNOVAE}

\author{A.M. Khokhlov}
\affil{Laboratory for Computational Physics and Fluid Dynamics, \\
Code 6404, Naval Research Laboratory,\\
Washington, DC 20375}

\author{E.S. Oran}
\affil{Laboratory for Computational Physics and Fluid Dynamics, \\
Code 6404, Naval Research Laboratory,\\
Washington, DC 20375}

\author{J.C. Wheeler}
\affil{Astronomy Department, The University of Texas\\
 Austin, TX 78712-1083}

\begin{abstract}
We derive the criteria for deflagration to detonation transition (DDT) 
in a Type Ia supernova. The theory is based on the two major assumptions:
{\it (i)} detonation is triggered  via the Zeldovich gradient mechanism
inside a region of mixed fuel and
products,  {\it (ii)} the mixed region is produced by a turbulent mixing of 
fuel and products either inside an active deflagration front or during the 
global expansion and subsequent contraction of an exploding white dwarf.

We determine the critical size of the mixed region required to initiate a  detonation
in a degenerate  carbon-oxygen mixture. This critical length is much larger than the
width of the reaction front of a Chapman-Jouguet detonation. However, at densities
greater than $\simeq 5\times 10^6$  \gcc, it is much smaller than  the size of a white
dwarf. We derive the critical turbulent intensity  required to create the mixed region
inside an active deflagration front  in which a detonation can form. We conclude that
the density $\rho_{\rm tr}$ at which a detonation can form in a carbon-oxygen white
dwarf is low, less than $2-5 \times 10^7$ \gcc, but  greater than $5\times 10^6$
\gcc. 

\end{abstract}


\section{Introduction}

The standard model for Type Ia supernovae (SN~Ia) is an explosion by a process of
thermonuclear combustion.  One-dimensional  calculations show that the two
possible modes of this combustion are either supersonic detonation or subsonic
deflagration (\cite{arn69}; Hansen \& Wheeler 1969; \cite{iic74}; \cite{nsn76}; \cite{nty84}).
There is evidence, both theoretical and empirical, that SN~Ia involve a transition
from a deflagration to a detonation, or DDT (Khokhlov 1991a,b;
\cite{al94};
H\"oflich, Khokhlov \& Wheeler 1995; \cite{hof95}; \cite{whkh95}; \cite{hk96} and references
therein). Pure detonation models tend to produce an excess of iron peak elements,
rather than the intermediate mass elements that are observed in the outer layers
of SN~Ia.  Models in which DDT occurs  can account for the properties of SN~Ia,
including the observed correlated dispersion in light curve  amplitude and decay
rate (\cite{hkw95}; \cite{hk96}).  In such models, the burning front starts as  a deflagration
that makes a transition to a detonation directly (the standard ``delayed detonation'' model),
or after the star has expanded due to partial burning and then recompressed (the
``pulsation delayed  detonation"). In these models, the density at which DDT
occurs is a free parameter.  The comparison of these models with observations
suggests that DDT must occur at relatively low densities, of the order of $10^7$ \gcc.
In this paper we address some of the key physical issues of DDT. In particular, we
address the physical mechanism by which DDT may occur in supernovae, and why it
might occur at low densities.  The process of DDT is not well understood in either
supernovae or in terrestrial combustion. Many of the principles presented here
apply to terrestrial conditions as well.

The analysis we present is based on a theory of unconfined DDT (Khokhlov,
Oran \& Wheeler 1996b). It is assumed that the inherent mechanism of DDT
is the Zeldovich  gradient mechanism (\cite{zlms70}) in which ignition
in a region of induction time gradient  leads to the development of a
reaction zone and a shock that coalesce to form a detonation. In this
theory, it is presumed that  a mixed region with gradients of temperature
{\sl and} composition, a combination of which results in an induction
time  gradient, is created  within a deflagration region by turbulence
that microscopically mixes  hot products and cold fuel. Two quantities
must be determined: the size of the critical region, L$_c$,  required for
a self-propagating detonation to develop,  and the amount of mixing (the
intensity or velocity of turbulence) required to produce that critical
length of mixing.

In supernovae, there are two situations in which DDT could occur. First, DDT
could occur directly, when turbulent deflagration alone creates a large mixed
region which ignites. For this process to occur, the flame surface  associated
with  laminar burning has to be destroyed by turbulence.  When the laminar
flame front is destroyed, reactions will still continue at rates depending on
the density and temperature, but microscopic chemical and thermal mixing of
fuel and product occurs.     The second way to initiate DDT is for the
deflagration to die due to expansion before the DDT occurs.  If insufficient
energy  has been liberated, the star will expand and then contract. Mixing of
fuel and products, and the subsequent DDT may
occur during the contraction phase.

In this paper, we first examine the mixing  process and determine the
thermodynamic properties of the mixture as a function of the degree of
mixing.  We show that these  thermodynamic properties can be
characterized by a single parameter (e.g., the mass fraction of fuel)
and  that the history of mixing, that is, whether mixing occurs before,
after,  or during a pulsation phase does not change the thermodynamic
properties of the mixture significantly.  We derive the critical length
scale to initiate a detonation by the Zeldovich mechanism. Next, we
evaluate the turbulent intensity necessary to break a laminar flame and to create the
conditions for DDT. This required the calculation of the laminar flame speed and 
laminar flame thickness over a wide range of densities, $3 \times 10^7 - 10^{10}$
\gcc. Finally, we estimate the critical density for DDT based on the existing
three-dimensional  simulations of turbulent deflagration in Type~Ia supernovae,
 and discuss implications of the
results for  supernova models.

\section{Thermodynamics of mixing}

Consider a subsonic mixing process that occurs at approximately constant pressure,
and assume that no reactions occur during mixing.  The conservation of mass and
enthalpy during mixing gives the  thermodynamic parameters of the mixture,
\begin{equation}
h(T_m,\rho_m) = X_f \, h(T_f,\rho_f)+(1-X_f) \, h(T_p,\rho_p)~,
\end{equation}
\begin{equation}
\frac{1}{\rho_m}=\frac{X_f}{\rho_f}+\frac{(1-X_f)}{\rho_p}~,
\end{equation}
\noindent
where $h$ is the enthalpy;  subscripts {\sl f}, {\sl p}, and {\sl m} stand for
fuel, product, and mixture of products and fuel, respectively; $X_f$ is the fuel
mass fraction in the mixture. We assume that the state of the products, {\sl p},
is determined by conservation of enthalpy during constant pressure burning in a
deflagration wave,
\begin{equation}
h(T_p,\rho_p)=h(T_f,\rho_f)+\Delta q~,
\end{equation}
\begin{equation}
P(T_p,\rho_p)=P(T_f,\rho_f)={\rm const}~,
\end{equation}
\noindent
where $\Delta q$ is the energy released by nuclear reactions.

The value of $\Delta q$ depends on the density at which the laminar flame is
either quenched by turbulence or quenched by expansion, and is not known exactly.
If the final stage of deflagration takes place at very low  densities, say, less
than $10^7 \gcc$, only carbon would be able to react  in a deflagration wave.
Temperatures resulting from carbon burning would be so low that oxygen would not
react. In this case, we expect $\Delta q \simeq 3.7\times 10^{17}$ ergs g$^{-1}$, and the
products of burning would consist predominantly of half silicon and half oxygen by
mass. At higher densities, both carbon and oxygen will burn to form silicon group
elements. For the latter case, we pick a typical value of  $\Delta q \simeq
6\times 10^{17}$ ergs g$^{-1}$. Below we refer to these two choices as Case A and B,
respectively.

Now consider mixing and changes in pressure due to global  adiabatic expansion or
contraction.  For a perfect gas, the order of  mixing and expansion or contraction
does not change the final state of mixed material.   For the equation of state of
white dwarf material, this is not an exact statement, but we show that it is a
good approximation.

Figure~1 shows the results of the following numerical experiment using the
degenerate matter equation of state described in the next section. A half carbon,
half oxygen  fuel at density $5\times10^6$ g cm$^{-3}$ (point~$f$ in Figure~1) was
burned at constant pressure to a half silicon and half oxygen mixture (to reach
point $p$) and then mixed with $X_f=0.5$  at constant pressure to yield the
conditions at point $m_o$.   Alternatively, both fuel and products were expanded
adiabatically  by a factor of 100 (to points $f^\prime,~p^\prime$), and then mixed
at constant pressure to produce the conditions at point $m^\prime$.   Finally, the
mixture was returned along the adiabat to the initial pressure.  The final state,
point $m$, is  very close to $m_o$, even under this extreme adiabatic expansion
and contraction.  The difference in temperature  between points $m$ and $m^\prime$
is 5\%, and the difference in density is 7\%. We conclude that, to a very good
approximation, the thermodynamic properties of a mixture do not depend on the
history of  the mixing and expansion processes.  We make that assumption
in the following analysis and neglect the order in which mixing and expansion or
contraction occur. 

Mixing cold fuel and hot products raises the entropy of  the mixture, so that
the mixture will ignite at lower density than the fuel alone.   For cases A and
B, Figure~2 gives the density at which the energy-release rate in the
fuel-product mixture is one-tenth of the maximum energy-release rate for
isochoric burning at this density. The maximum release rate occurs after
ignition, as the temperature rises and before fuel is consumed. Once
the mixture is at this density, the induction time is very short, only ten
times the characteristic time required to burn all of the fuel. The figure
shows that for a pure fuel to ignite, it needs to be compressed to very high
densities. However, even for fuel mixed with rather small amounts of hot
products, the ignition density decreases drastically, and is $\simeq 10^6 - 10^7$
 g
cm$^{-3}$ for a wide range in fuel fraction. If matter is returning to high
densities after a pulsation, a mixture containing  a noticeable amount of
products cannot be compressed to densities significantly in excess of 10$^7$ g
cm$^{-3}$
 before it ignites.  For the pulsation delayed detonation models, therefore, the thermodynamic
considerations place a limit on the conditions at which DDT can occur: DDT can only occur at
densities near or below 10$^7$ g cm$^{-3}$.  For the standard delayed detonation models, however,
no such limit can be placed by thermodynamic considerations. For both cases, the
kinetics and hydrodynamics of ignition need to be considered to determine whether
DDT can occur at thermodynamically allowed conditions.

\section{Critical length scale for DDT}

We now determine the critical size of the mixed region capable of triggering a
detonation.  It has been shown (\cite{kow96b}) that there are two processes that
are necessary for DDT to occur by the Zeldovich mechanism. First, a spontaneous
wave  must be generated and a shock-reaction complex must be formed.  This complex
forms somewhere in  the middle of the mixed region where there is both enough fuel
and high enough temperature. Second, this complex must be able to survive the 
propagation into the unburned material.

To estimate the critical size of this region, L$_c$, we  consider, as in
(\cite{kow96b}),  a nonuniform region created by mixing the products of isobaric
burning and fresh fuel, such that there is  a linear spatial distribution of fuel
fraction $X_f$,
\begin{equation}
X_f(x) = \cases{  x/L ,& $0 \le x \le L$  \cr
                1   ,& $x > L$ \cr}~,
\end{equation}
where $L$ is the size of the mixed region.  This creates a region of oppositely
directed  gradients in temperature and concentration of reactants. Initially, the
velocity of the material is zero, and the pressure $P_0$ is constant everywhere. 
The boundary conditions at $x=0$ are reflecting walls (symmetry conditions).

The system is described by the Euler equations coupled with the
nuclear kinetic equations
\begin{equation}
  {\partial\rho\over\partial t}~+~\nabla\cdot\left(\rho\vec{U}\right)~=~0~,   
\end{equation}
\begin{equation}
  {\partial\rho\vec{U}\over\partial t}
  ~+~\nabla\cdot\left(\rho\vec{U}\vec{U}\right)~+~\nabla P~=~0~,     
\end{equation}
\begin{equation}
  {\partial E \over \partial t} ~+~
  \nabla\cdot\left( \, \left( E+P \right) \vec{U} \, \right)
                  ~=~0~,
\end{equation}
\begin{equation}
  {\partial \cvec{Y} \over \partial t} 
   ~+~\left(\vec{U}\cdot\nabla\right) \cvec{Y}~=~ \cvec{R}~,     
\end{equation}
where the total energy density $E = E_t + \rho U^2/2 + 
\rho N_a \cvec{Q}\cdot\cvec{Y}$ is the sum of the thermal, kinetic, and 
nuclear energies,
$\vec{U}$ is the fluid velocity, $E_t$ is the
thermal energy density, $\cvec{Y}$ is the vector in the composition space
of mole fractions of nuclei,
$\cvec{Q}$ is the vector of corresponding binding energies per nucleon, and 
$\cvec{R}(\rho,E_t,\cvec{Y})$ is
the vector of net rates of change of nuclear species with time; $N_a$ is the
Avogadro number. To describe thermonuclear burning, an alpha-nuclei reaction
network is used with $\cvec{Y} = (\He, \C, \O, \Ne, \Mg, \Si, ^{32}S, \Ar, \Ca, \Ti,
\Cr, \Fe, \Ni)$. The equation of state includes contributions from Fermi gases of
electrons and positrons, radiation, and ions. Further details of the input physics
can be found in (\cite{kho93}).

The system of equations  is integrated numerically using a one-dimensional version
of the time-dependent, compressible fluid code ALLA based on the Piecewise Parabolic
Method (\cite{cw84}; \cite{cg85}), implemented as lagrangian step plus remap.
Details of the implementation are given in (\cite{kho95}; \cite{kow96a}).  The
nuclear reactions are coupled to fluid dynamics by time-step splitting.  The
kinetic equations are integrated together with the equation of energy conservation
using a stiff solver  with adjustable substeps to keep the accuracy better than
1\%. Most computations were done using a grid with 1024 equidistant
computational cells. The convergence of the numerical solutions was tested for
selected cases by varying the number of  cells from 1024 to 4096. Computations
were carried out in both planar and spherical geometries.

Computations were performed assuming that the products of burning that  formed the
mixture (state $p$ in the thermodynamic relations given in equations 1 to 4) are
characterized by the value of the energy released by the deflagration $\Delta q =
4\times 10^{17}$ ergs g$^{-1}$, or $\Delta q = 3\times 10^{17}$ ergs g$^{-1}$. These values are
close to Case A of \S\ 2 and have been selected to test the sensitivity of the results
to $\Delta q$. For $\Delta q = 4\times 10^{17}$, the composition of products was
taken to be pure \Si. For the second value, the  composition of products was taken
to be half \Si~ and half \O~ by mass.  The system was prepared in an initial state 
and then allowed to evolve in time, first until ignition takes place, then until
the formation (or failure) of detonation,  and finally to the time when the
generated detonation or shock leaves the computational domain. The results of these
computations are not sensitive to the assumed energy release and composition. They
mainly depend on the density at which  explosion occurs, and on the size of the
explosion region.

Figure 3 is an example of a computation in which a detonation successfully forms.
The initial density of the fuel $\rho_0 = 1.0\times 10^7$ \gcc and the size of the
mixed region is $L=5.0\times 10^5$ cm. Figures 3a through 3e show profiles of
density, temperature, pressure, velocity, and carbon mole fraction,  respectively,
for selected times. In the initial state, the density and carbon mole fraction of
the mixture  increase, and the temperature decreases as a function of position
$x$.  Because of such opposing temperature and  concentration gradients, the
ignition does not occur first at the origin, $x=0$, but is shifted to some point
$x > 0$ where the energy generation rate is maximum. From this point, the ignition
spreads out supersonically as a spontaneous wave. A detailed discussion of
spontaneous burning is given in Khokhlov et al. (1996b).  The overpressure and material
velocity  in the spontaneous wave grow  with time as the phase velocity of the burning front
decreases. When the speed of the spontaneous wave approaches the speed of sound, a shock wave
emerges, and a shock wave -- reaction complex forms. The profiles 1 to 3
illustrate the spontaneous wave propagation. There is no shock present at this
stage. Matter is  continuously compressed and accelerated in the process of
burning, and  expands and slows down afterwards. The shock -- reaction complex
forms  between the profiles 3 and 4. The complex continues to propagate through 
the matter in which density increases and temperature decreases  (profiles 4 and
5), but the strength of the complex grows.  Finally, the complex passes into the
fully unburnt cold fuel in the form of a slightly overdriven detonation (profile
6), and begins its relaxation to a Chapman-Jouguet state. 

Figures 3f and 3g show the nuclear composition for the times
corresponding to profiles 2 and 7, respectively. These figures illustrate
that at densities $\lta 10^7 \gcc$, the important reactions are
\C + \C $\rightarrow$ \Ne, \Mg, \He, and the reactions on \Ne and \Mg that 
transform the products of \C + \C burning into the \Si-group elements.
Reactions involving oxygen are less important. The initial amount of oxygen in the
mixture remains practically intact. The energy release is approximately $\Delta q 
\simeq (3.5-4) \times 10^{17}$ ergs g$^{-1}$, consistent with the  value of the 
energy release for the products of deflagration adopted above, and with 
the assumptions about the energy release made in \S\ 2. Figures 4a and 4b show
a successful initiation of a detonation at a density five times higher, 
$\rho_0=5\times 10^7$ \gcc. At this density, burning leads to higher  temperatures
due to a decrease in the specific heat of degenerate matter. As a result, oxygen
burns to silicon rather quickly (Fig.4b), and the energy of the explosion, 
$\Delta q\simeq 7\times 10^17$ ergs~g$^{-1}$, is 
almost two times larger than in the previous case
$\rho_0=10^7 \gcc$.

The shock--reaction complex can survive propagation down the temperature
gradient and can grow into a fully developed detonation only if  conditions in
front of the complex change slowly enough,  that is, if $L$ is large enough. If
the variations of preshock density and temperature with  distance are too fast,
the reaction zone structure is unable to adjust to the decreasing reaction rates
behind the shock, and is unable to compensate  for the energy required to shock
denser fuel.

Figures 5a to 5f illustrate the failure of the initiation of a detonation at
$\rho_0 = 10^7 \gcc$ when the size of the mixed region was 20 percent smaller,
$L=4\times 10^5$ cm, than in the case presented in Fig. 3. Here, too, the
spontaneous wave, and the shock -- reaction complex form (profiles 1 to 3).
However, the complex does not survive the propagation down the temperature
gradient. The shock and reaction  separate. The shock passes into the unburned
matter leaving the reaction wave behind. It is the condition that the complex must
survive the propagation down the temperature gradient,  not the formation of the
complex alone, that determines the critical size of the mixed region.

Figure 6 summarizes the results of a series of many computations performed to
determine $L_c$, the critical size of the region capable of triggering a
detonation, as a function of  fuel density.  The value of $L_{c}$ is a very
sensitive function of density. It is virtually independent of assumptions made
about the energy release $\Delta q$ and composition of the products formed during 
the preceeding deflagration stage. The critical length depends, however, on the
energy release during the explosion itself. As was mentioned earlier,  the energy
release increases with increasing background  density,
 which makes the dependence of $L_{c}$ on density very strong. The values of $L_c$
obtained in spherical and planar geometries are practically the same, because the
radius of  curvature of a spherical front is approximately the same as the
critical length which is much larger than the detonation wave thickness. 

The simulations show that the spontaneous wave grows to an appreciable strength,
and a shock -- reaction complex forms in a part of the mixed region with a high
fuel fraction, $X_f \simeq 0.8-0.9$. Thus, it is this mixture that is important
for DDT. As shown in Figure 2,  explosion of such a mixture can take place at
densities $\le 10^7 \gcc$ (Fig. 2). On the other hand, as shown in
Figure 6,  the critical length for DDT is a sensitive  function of density; it
increases steeply at low densities and it is of order $3\times 10^7-4\times 10^5$ cm at 
densities of $5\times10^6-10^7$ g cm$^{-3}$.  This length is a factor of
$10^4-10^5$ larger than the size of a detonation reaction zone.   However, even
at these low densities,  this length is small compared to the radius of  a
white dwarf. The length scale $L_c$ is small enough to be formed by  turbulent
mixing in a relatively small portion of the supernova ejecta for
$\rho\gta5\times10^6$ g cm$^{-3}$.
 
\section{Conditions on turbulent intensity}

Here we discuss the conditions necessary to produce a mixed region of size L$_c$ for
both the direct and the pulsation modes of delayed detonation. First consider DDT in
the direct mode.  A fundamental assumption is that turbulence must destroy the flame 
front so that burned and unburned materials can mix microscopically.  To break the
flame front, the turbulence velocity $U_b$,  at a scale comparable to the thickness of
the laminar front, must exceed the normal speed of the laminar flame,
\begin{equation}
U_b \simeq K\ S_\ell ,
\end{equation}
where $S_\ell$ is the laminar flame speed, $\delta_\ell$ is the  laminar flame
thickness, and $K\sim 1$ is a coefficient that describes the  ability of the laminar
flame to survive in a turbulent environment (\cite{kow96b}). 

The mechanism for breaking the flame is flame stretching, which includes the effects
of strain and curvature. These effects depend on the properties of the particular
flame and on the spectrum of the turbulence.  For terrestrial flames, the coefficient
$K$ significantly exceeds unity according to both theory and experiments
(\cite{pct96}; \cite{rdd93}). For the thermonuclear flame, the value of $K$ is not
known exactly.  The analysis of the nonlinear stabilization of the flames in a
turbulent field suggests $K=8$ (\cite{kho95}). The numerical simulations of the flame
presented in (\cite{kho95}) are consistent with a large value of $K$. For the
subsequent analysis in this paper, we use the two values $K=1$ and $K=8$. 

Figure 7 shows the laminar flame speed $S_l$ and thickness $\delta_l$ for a 0.5C +
0.5O mixture computed for the range of densities $3\times 10^7 - 10^{10}$ \gcc, as
described in the appendix.  In the same figure, we show $S_l$ according to  Timmes and
Woosley (1992) who computed the laminar flame speed at high densities, $\rho \ge 2
\times 10^8$~\gcc.  Our results are in good agreement with theirs at these densities.
At low densities the flame speed decreases with density more rapidly than at high 
densities, and reaches a very low value of $S_l\simeq 3\times 10^4$ cm/s at
$\rho\simeq 3\times 10^7$ \gcc. The computed values of $S_\ell$ give $U_b$ assuming
$K=1$. The curve $U_b^8 = 8 S_l$ shows the required turbulent intensity if $K=8$.  

To relate $U_b$ to the intensity of turbulent motions on 
larger scales, additional assumptions are required about the spectrum of turbulent motions
inside the white dwarf during the explosion. The most favorable assumption for
DDT is that the turbulent spectrum is Kolmogorov.
By assuming  that a 
Kolmogorov spectrum is established, we ignore the time delay required to
establish the steady-state cascade of 
turbulent energy, and ignore freeze-out of turbulence
due to the expansion of the star.
Under the assumption of a Kolmogorov
spectrum, the speed of turbulent motions at the scale of the laminar front $\delta_l$
is related to the speed of turbulent motions on larger scales
 $L > \delta_l$ as
\begin{equation}
U_{\delta_l}= U_L \left(\frac{\delta_\ell}{L}\right)^{1/3}~.
\end{equation}

During the deflagration phase of the explosion, large portions of the star will be
occupied by the very convoluted flame. To be specific, we pick a typical value $L=L_f
\simeq 10^8$ cm for the maximum size of the turbulent flame region (\cite{kho95}). The
turbulent region may be somewhat smaller in the beginning of the explosion, and
somewhat larger at the end, but this value is of the correct order of magnitude. If we
assume that the deflagration is strong enough to unbind the star, we have to assume
that $L_f$ is large enough. The large-scale turbulence in a supernova is driven by
buoyancy. If we neglect  the effects of freeze-out due to expansion of the star, we
expect $U_{L_f}$ to be of order of the Rayleigh-Taylor characteristic velocity at  this
scale $U_{L_f} \simeq  U_{R-T} \simeq 0.5 \sqrt{gL_f}$ where $g$ is the effective
gravitational acceleration.  In a star in hydrostatic equilibrium, or close to it, the
value of $\sqrt{gL_f}$ does not exceed the  speed of sound. In  equilibrium, $ P/R
\sim \rho g$, where R is of order the stellar radius. From this it follows that $a_s
\simeq \sqrt{P/\rho} \sim \sqrt{gR}$ and hence $\sqrt{gL_f}<a_s$ for $L_f<R$. For  our
estimates we take the value of $U_{L_f} = U_{R-T} = 0.5a_s$. The values of the
turbulent velocity at the scale of the flame front estimated from equation (11) using
these values is plotted in Figure 7.  The intersection of the curve  $U_{\delta_l}$
with the curves $S_l$ and $U_b^8$ gives estimates of the transition density $\rho_{\rm
tr} \le (4-8)\times 10^7$ \gcc below which DDT may occur. We emphasize that this
estimate of $\rho_{\rm tr}$ is made neglecting any time delay in establishing the
Kolmogorov spectrum, and neglecting freeze-out of turbulence due to expansion of the
star.

In order to take the effect of freeze-out into account, three-dimensional
simulations of the explosion of the entire star are required. To date, only one
such simulation has been performed (\cite{kho95}), and the analysis presented
in that paper suggests that the expansion freezes scales exceeding approximately
$L\simeq 10^6-10^7$ cm, and that the turbulent burning velocity at these scales
does not exceed $U_L \simeq 10^6 - 10^7$ cm/s. If we take from this range the
 most favorable values for DDT, $U_L=10^7$ cm/s and  $L=10^6$ cm, equation (11)
gives another estimate of the turbulent velocity
  at the  scale of the flame front, $U_{\delta_l}^c$,
which is also shown in Figure 7. The intersection of the curve $U_{\delta_l}^c$ 
with the curves $U_b^8$ and $S_l$ gives, we believe, a more realistic
estimate of the transition density $\rho_{\rm tr} \le (2-5)\times 10^7$ \gcc.

There are other effects that may influence the transition density. Self-turbulization
of the flame on scales less than $L_f$ might be possible, as is preconditioning of a
mixture by shocks generated by the turbulence itself. The latter effects might
somewhat facilitate DDT, but not to a great extent (Khokhlov et al. 1996b). 
Self-turbulization due to the Landau-Darrieus  instability is limited by nonlinear
effects. Shock preconditioning requires a very  strong shock to be effective.  We
believe that there are two major unknown factors in estimating the value of $\rho_{\rm
tr}$: the  spectrum of turbulence in the exploding star, and the extent to which the
flame can survive turbulent stretch on small scales (coefficient $K$).

Now consider the case in which the conditions  for DDT do {\it not} occur in the
deflagration phase, and the deflagration is extinguished  by the expansion.  In this
case, burning is quenched by expansion and the products and fuel mix freely.   There 
is no laminar flame  front and there are no nuclear reactions.  This situation  is
still Rayleigh-Taylor unstable.  Previous estimates (\cite{kho91b}; \cite{al94}) have
shown that the mixed R-T region may be on the order  of $10^6-10^8$ cm, which is much
larger than $L_c$ at $\rho \gta 5\times10^6-10^7\gcc$. With a Kolmogorov cascade, the
eddy turnover time scales as $\lambda^{2/3}$, where $\lambda$ is the size of the eddy. 
When the star is compressed, the turbulent motions on all scales will be enhanced and
mixing should proceed vigorously (this is opposite to the effect of freeze-out
expected during the expansion of the star). It is thus reasonable to expect
microscopic mixing to be complete before the contraction phase.\footnote{We have
invoked microscopic mixing in this model.   There is considerable evidence that mixing
occurs in supernova ejecta but is not always microscopic.  Several lines of evidence
strongly suggest that the mixing in SN~1987A was  macroscopic, but not microscopic. 
We note, however, that the physical conditions here and in a collapse environment like
SN~1987A are different. The instability leading to mixing in SN~1987A is expected to
be a Richtmeyer-Meshkov instability in which the density/pressure inversion is only
temporarily induced at composition/density boundaries due to shock passage. In the
thermonuclear explosions that we consider here, the Rayleigh-Taylor instability is
intrinsic and long lasting. There should be ample time to mix to small length scales.}
This suggests that the pulsation mechanism can lead rather naturally to conditions
where DDT is unavoidable in the recompression phase.

\section{Conclusions}

We applied a theory of DDT in unconfined
conditions (\cite{kow96a}) to DDT in Type~Ia supernovae. The two basic assumptions
of this theory are: 1) the gradient mechanism is the inherent mechanism that
leads to DDT in unconfined conditions, and 2) the mechanism for preparing the
gradient of induction time is turbulent mixing that requires breaking or quenching the
flame front.

Using a series of numerical simulations, we determined the minimum critical size
of the mixed region, $L_c$, required for DDT (\S\ 3). We find that $L_c$,
though much larger than the detonation wave thickness, is at least two to three
orders of magnitude smaller than a white dwarf radius for $\rho>5\times 10^6- 10^7$
 \gcc.
Thus, a mixed region of critical size
required for triggering a detonation can, in principle, be formed by turbulent 
mixing in a small portion of a  supernova.

There are two possible modes of detonation formation in a Type~Ia supernova:
a direct mode when DDT occurs inside an active deflagration front, and a
pulsation mode which is possible if DDT does not occur directly, and if
the energy released during the deflagration phase is not enough to unbind 
the star. Then the star will experience a global pulsation, and mixing
will take place between the cold fuel and ashes of the dead deflagration.
The requirements on the intensity of turbulent mixing needed to form
a mixed region of critical size $L_c$ are different in these two cases.

For a direct mode of DDT (delayed detonation model), the intensity of turbulence
on small scales of the order of the laminar flame thickness  must be
high enough to stretch and  break the surface of the laminar flame front (\S\ 4,
equation 10). We computed the laminar flame velocity and thickness
in a wide range of densities $3 \times 10^7 - 10^{10}$ \gcc, and estimated the
 critical intensity of turbulence required to break the flame at the
 scale of a laminar flame thickness.

We also tried to estimate the value of the transition density
$\rho_{\rm tr}$ at which DDT should occur during a Type~Ia supernova explosion. Using
available information about the intensity of turbulence inside the exploding star, we
conclude that a realistic estimate of the transition density for the direct delayed
detonation explosion is $\rho_{\rm tr} \lta (2-5)\times 10^7 \gcc$. However, even under
the most favorable and unrealistic  assumptions (\S\ 4), $\rho_{\rm tr}$ does not
exceed $10^8$ \gcc.

In the pulsation mode (pulsating delayed detonation model), microscopic mixing
throughout large areas can occur because there is no flame front separating burned and
unburned matter, and, thus, fuel and products can mix freely. We considered the
thermodynamics of the mixing process in a supernova (\S\ 2) to  demonstrate that the
order of mixing and global expansion or contraction of the supernova is not
important.  We also found that even a small fraction ($\gta 1\%$) of  high-entropy
products of burning mixed with cold, low-entropy fuel increases the entropy of the
mixture such that it will ignite at very low densities. We conclude that in this mode, 
the portions of the  mixture of size comparable to or exceeding $L_c$ would
unavoidably detonate around $\rho_{\rm tr}\simeq 10^7~\gcc$.

Despite various details of DDT that require future consideration, our main conclusion
is that DDT in Type~Ia supernovae should occur  in the density range $5 \times
10^6$ to a few times $10^7$. At higher densities, $10^8-10^9$ \gcc, either the required
turbulent velocity cannot be reached  (direct mode of the  delayed detonation) or the
material inevitably ignites before such high densities are reached (pulsating mode of
the delayed detonation). At lower densities, the size of the critical region becomes too
large. This conclusion agrees with the results of global modeling of Type~Ia
supernovae, where the transition density, $\rho_{\rm tr}$, at which DDT occurs is a
free parameter. Global modeling suggests that, in order to fit observed light curves
and spectra of SNIa,  the transition density should be $\rho_{\rm tr}\simeq 10^7~\gcc$
(\cite{kho91a}; \cite{ww94}; \cite{hkw95}; \cite{nom95}; \cite{hof95}; \cite{hk96}).

\acknowledgments
This work was supported by  NSF grants 92-18305 and 94-17083, NASA grant NAG52888, 
Texas Advanced Research Program grant ARP 469, and by the Naval Research Laboratory 
through the Office of Naval Research. We thank P.A. H\"oflich for many important
discussions and F.-K.\ Thielemann for providing the thermonuclear reaction rates.
Input for the electron-conductivity calculations was kindly provided by  D.\
Yakovlev. Peter Hauschildt and the anonymous referee made very useful comments that
led to improvements of the paper during the refereeing process.

\clearpage
\appendix
\section{Properties of laminar flames in degenerate C-O mixtures}

The velocity of a laminar flame $S_l$ is computed as 
the eigenvalue of the differential equations describing the
structure of a steady-state burning wave (\cite{zblm85}). 
We assume that the wave propagates steadily and that burning occurs at
constant pressure since the flame speed is much less than the sound speed.
The parameters of the fuel and the products of burning are designated below by
subscripts 0 and 1, respectively.
The mass flux is constant through a steady wave, $J_m = \rho u = \rho_0 S_l
=\rho_1 u_1$.
The equation of energy balance inside the wave is
$$
J_m~c_P~{dT \over dx}~=~
{d \over dx} \left( \chi {dT \over dx} \right)
 ~+~ J_mN_a\cvec{Q} \cdot \cvec{R}~,    \eqno(A.1)
$$
where $c_P$ is the specific heat at constant pressure, $\chi$ is the
thermal conductivity, $N_a$ is the Avogadro number, 
$\cvec{Q}$ is the vector of binding energies of nuclei,
$\cvec{R}$ is the vector of net reaction rates (Section 3), and
$x$ is the distance across the wave. The terms on the
right side of (A.1) describe the temperature changes due to
the heat conduction and nuclear reactions.
Nuclear mole fractions $\cvec{Y}$ are changing inside the wave due to
nuclear reactions,
$$
J_m \, {d \cvec{Y} \over dx} = \rho\, \cvec{R}~.   \eqno(A.2)
$$
Ion diffusion is ignored in (A.2) since, in a degenerate gas, the
mean free paths of ions are much smaller than the mean free paths of
electrons and photons.
The density $\rho(T,\cvec{Y})$ inside the wave
 is related to $T$ and $\cvec{Y}$ via the constant pressure condition
$$
     P(\rho,T,\cvec{Y}) = P_0~.                      \eqno(A.3)
$$
The boundary conditions for Eqs.(A.1) to (A.3) are
$$
{\rm at}~x=- \infty:
~~T = T_0,~\rho = \rho_0,~\cvec{Y} = \cvec{Y}_0,~{dT\over dx}=0,
                                                             \eqno(A.4)
$$
and
$$
{\rm at}~x = \infty:~~T = T_1,~\rho = \rho_1,~\cvec{Y} = \cvec{Y}_1,
       ~{dT\over dx}=0~.                        \eqno(A.5)
$$
The points $x = \pm \infty$ are singular points of equations (A.1)--(A.3).
The integral curve  which passes through these singular points
represents the solution for a steady flame front.
We designate the value of $J_m$ corresponding to this integral curve as $J^*$.

For low temperatures,  $T < T_{\varepsilon}$, where  $ T_{\varepsilon}$ is some value
less than $2 \times 10^9$K, we can neglect the  energy generation term
in (A.1) in comparison with the term
describing the heat transport. At these temperatures, (A.1) can be rewritten 
as
$$
{dW \over dT} ~=~ {J_m c_P\over \chi}          \eqno(A.7)
$$
where $W=\chi {dT\over dx}$. The corresponding initial condition for (A.7)
is $W(T_0) = 0$. 
The integration of (A.7) from $T_0$ to $T_{\varepsilon}$ gives a certain
value of $W_{\varepsilon} = W(T_{\varepsilon})$, from which we find the
corresponding value of $\left(dT\over dx\right)_{T_\varepsilon}$. From this 
point, we integrate the full equation (A.1) with the energy generation
term included.

The value $J^*$ is found using the method of trial and error 
(\cite{zblm85}).
The integral curves for $J_m \neq J^*$
diverge from the $J^*$ curve when
$x \rightarrow \infty$. For $J_m < J^*$ 
the asymptotic ($x \rightarrow \infty$) behavior is
$T \rightarrow - \infty$, ${dT\over dx} \rightarrow -\infty$.
For $J_m > J^*$ the asymptotic behavior is
$T \rightarrow  \infty$, ${dT\over dx} \rightarrow \infty$.
We use the difference in the behavior of the integral curves for
discriminating between 
the cases $J_m < J^*$ and $J_m > J^*$. The next trial value of $J_m$
is found as the average of the two most recent trial values lying on
different sides of $J^*$.
It requires $\simeq 20$ iterations to obtain the value of 
$J^*$ with an accuracy of $10^{-3}$. Then the laminar flame speed is calculated as
$S_l = J^* / \rho_0$. The values of $\delta_l$ are estimated
approximately as the thickness of a zone 
where most of the nuclear energy is released
and the CO mixture is converted into Si-peak elements.
Computations are done for the initial mixture of equal masses of
$^{12}C$ and  $^{16}O$ with a small amount of
$^{22}Ne$ added in order to get a neutron excess $\eta = 0.002$.
The initial temperature is $T_0=10^8$K, and we  used  $T_{\varepsilon} = 10^9$ K.
Flame parameters are not sensitive to
variations of $T_0$ and $T_\varepsilon$ if these temperatures are
below $2\times 10^9$K.

In computations of $S_l$ and $\delta_l$,
the following 133 species reaction network is used:
$n$, $p$, $^4He$, $^{12,13}C$, $^{13}N$,
$^{16}O$, $^{20,22}Ne$,
$^{23}Na$, $^{23-26}Mg$, $^{27}Al$, $^{27-32}Si$, $^{30-33}P$,
$^{31-36}S$, $^{35-37}Cl$,
$^{36-41}Ar$, $^{39-43}K$, $^{40-46}Ca$, $^{43-47}Sc$, $^{44-50}Ti$,
$^{47-52}V$, $^{48-56}Cr$, $^{51-60}Mn$, $^{52-62}Fe$,
$^{55-61}Co$, $^{56-64}Ni$, $^{57-65}Cu$, $^{59-66}Zn$.
The forward reaction rates are taken from Thielemann's 
reaction rate library (\cite{tat87}).
The binding energies and partition functions were taken according to
Woosley et al. (1987). The backward
reaction rates are calculated according to the principle of
detailed balance.
Screening corrections are applied to the most important reactions,
$^{12}C + ^{12}C$, $^{12}C + ^{16}O$, $^{16}O + ^{16}O$ and
$3~^{4}He \rightarrow ^{12}C$, which control the thermonuclear energy
release at low temperatures. Even for these reactions screening
factors were found to be relatively small ($E \la 2 - 5$) for conditions
where nuclear reactions significantly contribute to the right side
of Eq.(2). Most of the nuclear reactions occur at high temperatures
and the corresponding screening factors $E \la 1.5$ are small
in comparison with the expected
uncertainties in the nuclear reaction rates ($\sim 2-3$).
Screening corrections are taken from Yakovlev and
Shalybkov (1989).
 
The total thermal conductivity is calculated as the sum of electron
and photon conductivities
$$
\chi ~=~ \chi_e ~+~ \chi_{\gamma},                        \eqno(A.8)
$$
with $\chi_{\gamma} = 4acT^3 / 3 \kappa \rho$. The expression for the
photon opacity (\cite{pac83})
$$
\kappa~=~ { {0.4~Y_e} \over {( 1 + 2.7~10^{11} \rho / T^2 )}}~
         { \left( 1 + \left( T \over 4.5~10^8 \right)^{0.86} \right)}^{-1}
          ~{\rm cm^2 \over g}                             \eqno(A.9)
$$
approximates the low density ($\la 6~10^7 \gcc$) results of Buchler and Yueh (1976)
that take the degeneracy of electrons and relativistic scattering into account.
$Y_e$ is the electron mole fraction.
At higher densities, the contribution of the photon
opacity to the total opacity is very small.
Our results at high densities are in a good agreement with
those of Timmes \& Woosley (1992) who used different approximations for the
photon thermal conductivity.
 
The electron thermal conductivity can be expressed
in terms of the effective electron collisional frequency, $\nu_e$, as
(Yakovlev and Urpin 1980)
$$
\chi_e  ~=~ 4.09~10^9 ~ T ~ { x^3 \over \sqrt{1+x^2} }
      \left( 10^{16} \over \nu_e \right)~{\rm ergs \over cm~s~K}
                                                    \eqno(A.10)
$$
where  $x = p_{\rm F} / m_e c
\simeq 1.009~10^{-2} ( \rho Y_e )^{1/3}$, $p_{\rm F}$ is the
electron Fermi momentum, and $\nu_e$ is a sum of ion-electron and 
electron-electron collisional frequencies,
$$
\nu_e ~=~ \nu_{ei} ~+~ \nu_{ee}.                    \eqno(A.11)
$$
The electron - ion collisional frequency is calculated using
the mean ion approximation following Urpin \& Yakovlev (1980)
$$
\nu_{ei} ~=~ 1.78~10^{16} \sqrt{1+x^2} ~ { Y Z^2 \Lambda \over Y_e },
                                                   \eqno(A.12)
$$
where $Y=\sum Y_i$ and $Z=Y_e/Y$ are the mean ion mole fraction
and the mean atomic number, respectively, and
$\Lambda$ is the
Coulomb logarithm which can be expressed for gaseous and liquid ionic
states
as the sum of the logarithm obtained in the Born approximation
and the non-Born correction (Yakovlev and Urpin 1980, Yakovlev 1987)
$$
\Lambda ~=~ \Lambda_{\rm B} ~+~ \delta \Lambda,\eqno(A.13)
$$
with
$$
\Lambda_{\rm B} ~=~ {\rm ln} ~ \left(
           { \left( { 2 \pi Z \over 3 } \right) }^{1/3}
            \sqrt{ 1.5 + {3 \over \Gamma } }
                         \right) ~-~ { x^2 \over 2(1+x^2) }~, \eqno(A.14)
$$
$$
\Gamma = 2.275~10^5 ~ { Z^{5/3} \over T } ~ { ( \rho Y_e ) }^{1/3}
                                                            \eqno(A.15)
$$
and
$$
\delta \Lambda ~=~ { \pi \over 2} \alpha \beta^2
       { 1 + 1.30 \alpha \over 1 + \alpha^2 ( 0.71 - 0.54 \beta^2) }~,
                                                       \eqno(A.16)
$$
$$
\beta = { x \over \sqrt{1+x^2} } ~~~,
         ~~~ \alpha = { Z \over 137 \beta} ~~~.    \eqno(A.17)
$$

The thermal conductivity due to electron-electron collisions
of relativistic electrons in a degenerate electron gas was calculated
in the static screening approximation
by Urpin and Yakovlev (1980),
$$
\nu_{ee} ~=~ 0.511 ~ T^2 ~ { x^{3/2} \over { (1+x^2) }^{5/4} } ~ J(y)
          ~ {\rm s^{-1}}~~~,
                              \eqno(A.18)
$$
where
$$
y ~=~ { \sqrt{3}T_{pe} \over T }~~~,\eqno(A.19)
$$
and
$$
T_{pe} ~=~ { \hbar \omega_{pe} \over k_{\rm B} }
       ~=~ 3.307~10^8 ~ { x^{3/2} \over { (1+x^2) }^{1/4} }~~{\rm K}
                                                     \eqno(A.20)
$$
is the electron plasma temperature. Recently Yakovlev (unpublished)
recalculated the integral
entering the general expression of $\nu_{ee}$ numerically and
obtained the following analytical expression for $J(y)$
$$
J(y) ~=~ {1 \over 3} ~ { \left( { y \over 1+ay} \right) }^3
         ~ {\rm ln} \left( { 2 \over y } + b \right)      \eqno(A.21)
$$
with $a=0.113$ and $b=1.247$. The largest error in the fit is $\sim 6$\%
at $y \simeq 2$.

\clearpage

Fig. 1 -- Comparison of thermodynamic properties of mixtures $m$ and 
$m_0$ formed by mixing the same products and fuel at high and at
low pressure, respectively.

Fig. 2 -- Ignition density as a function of the fuel fraction.
(Solid line) -- $\Delta q = 3.7\times10^{17}$ ergs g$^{-1}$;
(Dashed line) -- $\Delta q = 6.8\times10^{17}$ ergs g$^{-1}$.

Fig. 3 -- Initiation of a detonation for $\rho_0=10^7~\gcc$ and $L=10^5$ cm.
         Profiles are shown for the initial  time $t=0$, and 
          for times 
          $t= 2.86\times10^{-5}$ (1),
          $7.05\times10^{-5}$ (2),
          $1.40\times10^{-4}$ (3),
          $1.63\times10^{-4}$ (4),
          $1.90\times10^{-4}$ (5),
          $2.33\times10^{-4}$ (6),
          $2.76\times10^{-4}$ (7) and
          $3.27\times10^{-4}$ (8) sec.
(a) Pressure. (b) Temperature. (c) Density. (d) Velocity. (e) Carbon mole fraction.
(f) Distribution of nuclear species at the time corresponding to profile 2. (g)
Distribution of nuclear species at the time corresponding to profile 7.

Fig. 4 -- Initiation of a detonation for $\rho_0=5\times 10^7~\gcc$ and 
          $L=1.3\times 10^3$ cm.
          Profiles are shown for the initial  time $t=0$, and 
          for the moments of time 
          $t=1.74\times10^{-8}$ (1),
          $7.89\times10^{-8}$ (2),
          $1.38\times10^{-7}$ (3),
          $1.90\times10^{-7}$ (4),
          $2.41\times10^{-7}$ (5),
          $3.38\times10^{-7}$ (6),
          $4.31\times10^{-7}$ (7) and
          $4.96\times10^{-7}$ (8) sec.
(a) Pressure. (b) Distribution of nuclear species at the time corresponding to
profile 7. 

Fig. 5 -- Failed initiation of a detonation for $\rho_0=1\times 10^7~\gcc$ and 
          $L=8\times 10^4$ cm.
          Profiles are shown for the initial  time $t=0$, and 
          for the moments of time 
          $t=5.64\times10^{-5}$ (1),
          $1.12\times10^{-4}$ (2),
          $1.68\times10^{-4}$ (3),
          $2.22\times10^{-4}$ (4),
          $2.77\times10^{-4}$ (5) and
          $3.31\times10^{-4}$ (6) sec.
(a) Pressure. (b) Temperature. (c) Density. (d) Velocity. (e) Carbon mole fraction.
(f) Distribution of nuclear species at the time corresponding to profile 5. 

Fig. 6 -- Critical length for the initiation of a detonation as a function of 
the fuel density. 
(Circles) -- $\Delta q = 4\times 10^{17}$ ergs g$^{-1}$ for deflagration products.
(Triangles) -- $\Delta q = 3\times 10^{17}$ ergs g$^{-1}$ for deflagration products (all
but one triangle coincide with the circles).

Fig. 7 -- Velocities and flame thickness as a function of density. $S_l$ and
$\delta_l$ -- laminar flame speed and thickness, computed in the appendix. $S_l^{TW}$
-- laminar flame speed according to Timmes \& Woosley (1992). $U_b^8$ -- critical
turbulent velocity assuming $K=8$. $U_{\delta_l}$ -- estimated turbulent velocity at
the scale of the flame front during the supernova explosion, assuming no turbulence
freeze-out.  $U_{\delta_l}^c$ -- same, but  assuming turbulence freeze-out.  

\clearpage

\begin{thebibliography}{99}

\bibitem[Arnett 1969]{arn69} 
    Arnett, W.D., 1969. Ap\&SS, 5, 180.

\bibitem[Arnett \& Livne 1994]{al94}
    Arnett, W.D. \& Livne, E., 1994, ApJ, 427, 314.

\bibitem[Buchler \& Yueh 1976]{by76}
    Buchler J.R. \& Yueh W.R., 1976, ApJ, 210, 440.

\bibitem[Collela \& Woodward 1984]{cw84} 
   Colella, P., and Woodward, P.R., 1984. Journ.\ Comp.\ Phys., 54, 174.

\bibitem[Collela \& Glaz 1985]{cg85} 
   Colella, P., and Glaz, H.M., 1985. Journ.\ Comp.\ Phys., 59, 264.

\bibitem[Hansen, Torrie \& Vieillefosse 1977]{htv77}
   Hansen J.P., Torrie G.M. \& Vieillefosse P., 1977, Phys. Rev., A16, 2153.


\bibitem[Hansen \& Wheeler]{hw69}    
   Hansen, C.J. \& Wheeler, J.C., 1969. Ap\&SS, 3, 646.

\bibitem[H\"oflich 1995]{hof95}    
   H\"oflich, P., 1995. ApJ, 443, 89.

\bibitem[H\"oflich et al.\ 1995]{hkw95}    
   H\"oflich, P., Khokhlov, A.M., \&  Wheeler, J.C., 1995, ApJ, 444, 831

\bibitem[H\"oflich  \& Khokhlov 1996]{hk96}    
   H\"oflich, P. \& Khokhlov, A.M., 1996. ApJ, 457, 500.

\bibitem[Ivanova et al.\ 1974]{iic74}    
   Ivanova,  Imshennik, V.S. \& Chechetkin, V.M., 1974, Ap\&SS, 31, 497.

\bibitem[Khokhlov 1991a]{kho91a}  
   Khokhlov, A.M. 1991a, A\&A, 245, 114.

\bibitem[Khokhlov 1991a]{kho91b} 
   Khokhlov, A.M. 1991b, A\&A, 245, L25.
           
\bibitem[Khokhlov 1993]{kho93}    
   Khokhlov, A.M., 1993. ApJ, 419, 200.

\bibitem[Khokhlov 1995]{kho95}    
   Khokhlov, A.M., 1995. ApJ, 449, 695.

\bibitem[Khokhlov et al.\ 1996a]{kow96a}    
   Khokhlov, A.M. Oran, E.S. \& Wheeler, J.C., 1996a, Combustion \& Flame, 
    105, 28.

\bibitem[Khokhlov et al.\ 1996b]{kow96b}    
    Khokhlov, A.M. Oran, E.S. \& Wheeler, J.C., 1996b, Combustion \& Flame, 
    in press.

\bibitem[Nomoto et al.\ 1976]{nsn76}    
    Nomoto, K., Sugimoto, D. \& Neo, S., 1976. Ap\&SS, 39, L37.

\bibitem[Nomoto et al. 1984]{nty84}
    Nomoto, K., Thielemann, F.-K. \& Yokoi, K., 1984. ApJ, 286, 644.

\bibitem[Nomoto et al.\ 1995]{nom95}    
    Nomoto, K., Yamaoka, H., Shigeyama, T., Iwamoto, K., 1995.
    in {\it Supernovae and Supernova Remnants}, ed. R. McCray and Z. Wang,
    proc. of the IAU Colloquium 145 (Cambridge Univ. Press), p. 

\bibitem[Paczynski 1983]{pac83}
    Paczynski B., 1983, ApJ, 267, 315.

\bibitem[Poinsot, Candel \& Trouv\'e 1996]{pct96}
    Poinsot, T., Candel, S., \& Trouv\'e, A., 1996, Prog. Energy Combust. Sci.,
    21, 531.

\bibitem[Roberts et al. 1993]{rdd93}
    Roberts, W.L., Driscoll, J.F., Drake, M.C., \& Goss, L.P., 1993, Combust. Flame,
    94, 58.

\bibitem[Thielemann, Arnould \& Truran 1987]{tat87}
     Thielemann, F.-K., Arnould, M., Truran, J.W., 1987, {\sl Advances in Nulcear
     Astrophysics}, ed.\ E.\ Vangioni-Flam (Editions fronti\`eres: Gif sur Yvette),
     p.\ 525.

\bibitem[Timmes \& Woosley 1992]{tw92}
    Timmes, F. \& Woosley, S.E., 1992, ApJ, 396, 649.

\bibitem[Urpin \& Yakovlev 1980]{uy80}
    Urpin V.A. \& Yakovlev D.G., 1980, Sov. Astron.,  24, 126.


\bibitem[Wheeler et al.\ 1995]{whkh95}
    Wheeler, J.C, Harkness, R.P., Khokhlov, A.M. \& H\"oflich, P.A., 1995,
    Physics Reports, 256, 211.

\bibitem[Woosley et al. 1987]{whfz87}
    Woosley, S.E., Holmes, J.A., Fowler, W.A. and Zimmerman, B.A., 1987,
   Atomic Data and Nuclear Data Tables,  22, 371.


\bibitem[Woosley \& Weaver 1994]{ww94}
    Woosley, S.E. \& Weaver, T.A. 1994. in {\it Supernovae}, 
    ed. S. Bludman, R. Mochkovitch \& J. Zinn-Justin (Les Houches Vol. 24:
    Elsevier), p. 

\bibitem[Yakovlev 1987]{yak87}
    Yakovlev D.G., 1987,  Sov. Astron., 31, 347.


\bibitem[Yakovlev \& Shalybkov 1989]{ys89}
    Yakovlev D.G. \& Shalybkov D.A., 1989, Sov. Sci. Rev. E., 7, 311.

\bibitem[Yakovlev \& Urpin 1980]{yu80}
    Yakovlev D.G. \& Urpin V.A., 1980, Sov. Astron.,  24, 303.


\bibitem[Zeldovich et al.\ 1970]{zlms70}    
    Zeldovich, Ya.B., Librovich, V.B., Makhviladze, G.M. \& Sivashinsky, G.I,
    1970, Astronaut. Acta 15, 313.


\bibitem[Zeldovich et al. 1985]{zblm85}
    Zeldovich, Ya.B., Barenblatt, G.I., Librovich, V.B., Makhviladze, G.M., 1985,
    {\sl The Mathematical Theory of Combustion and Explosions} (Consultants Bureau: 
    London).

\end{thebibliography}
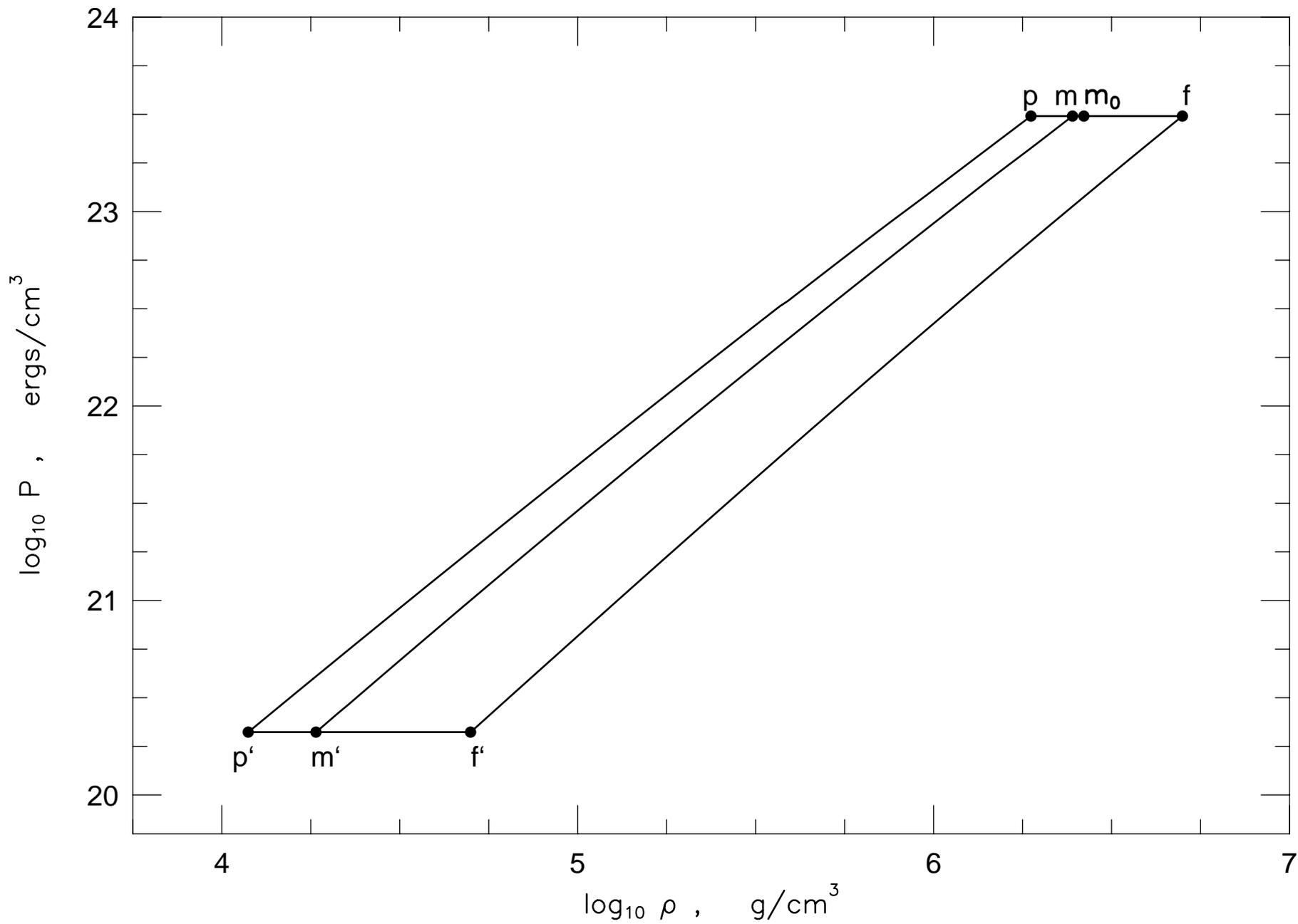
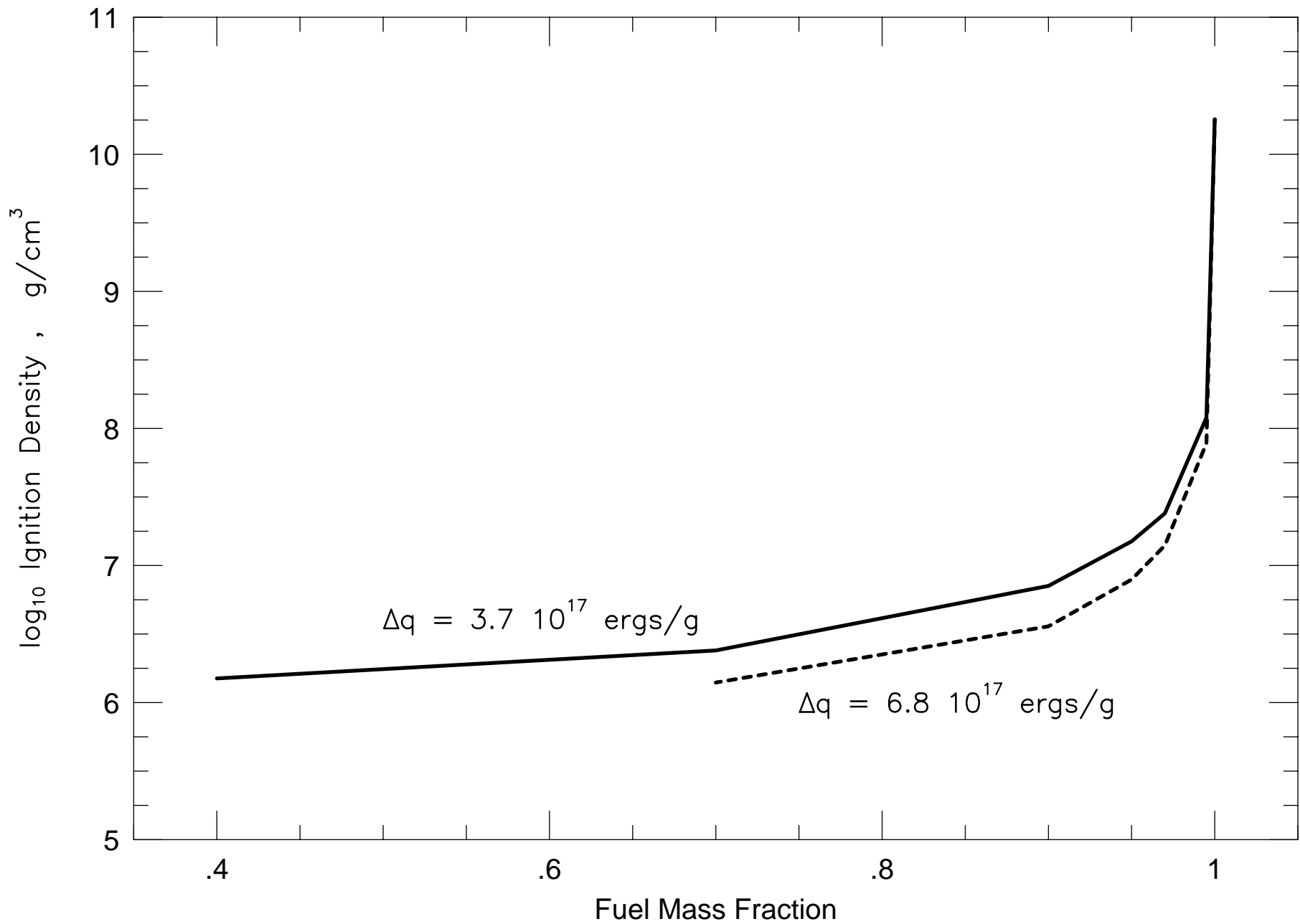
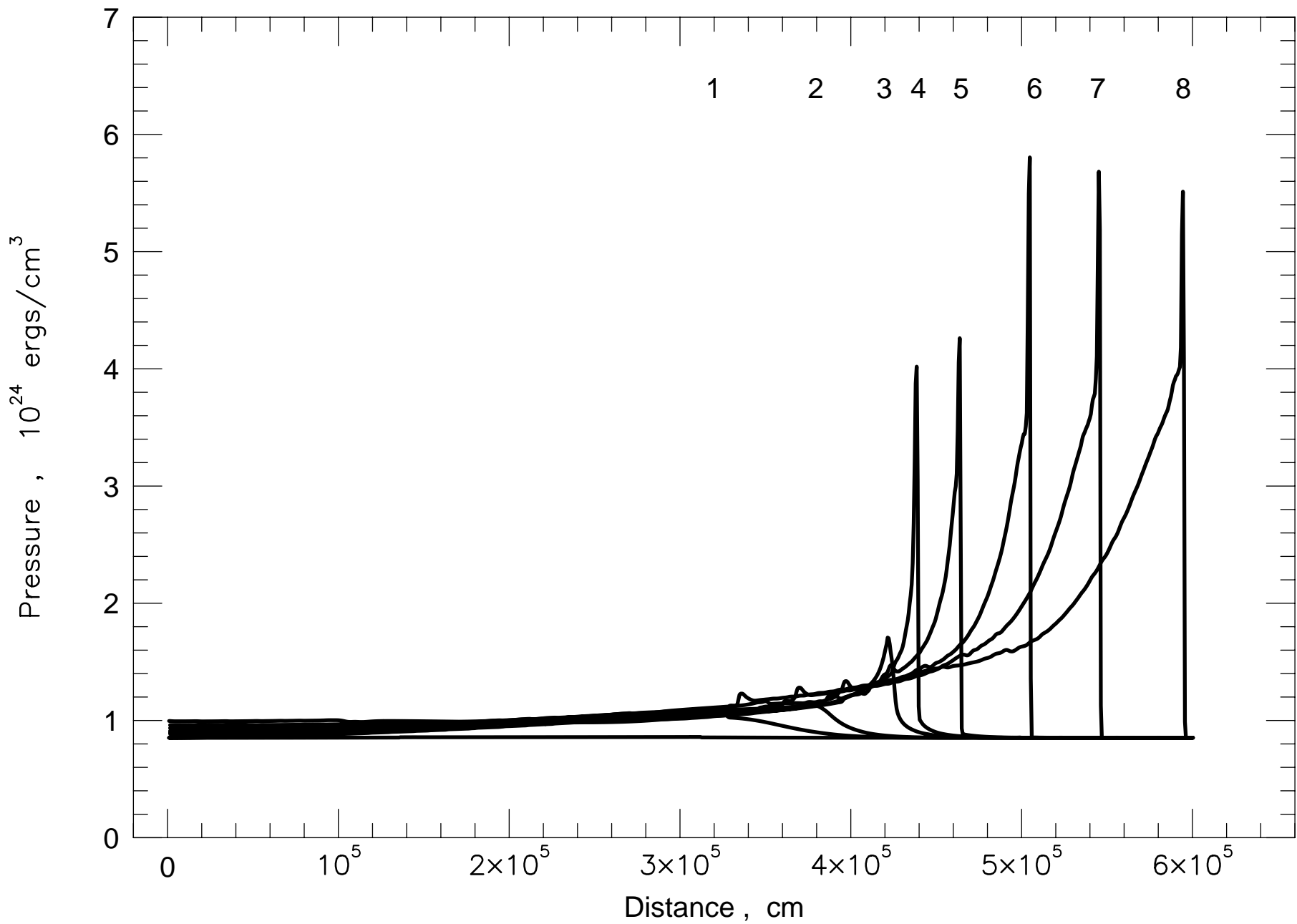
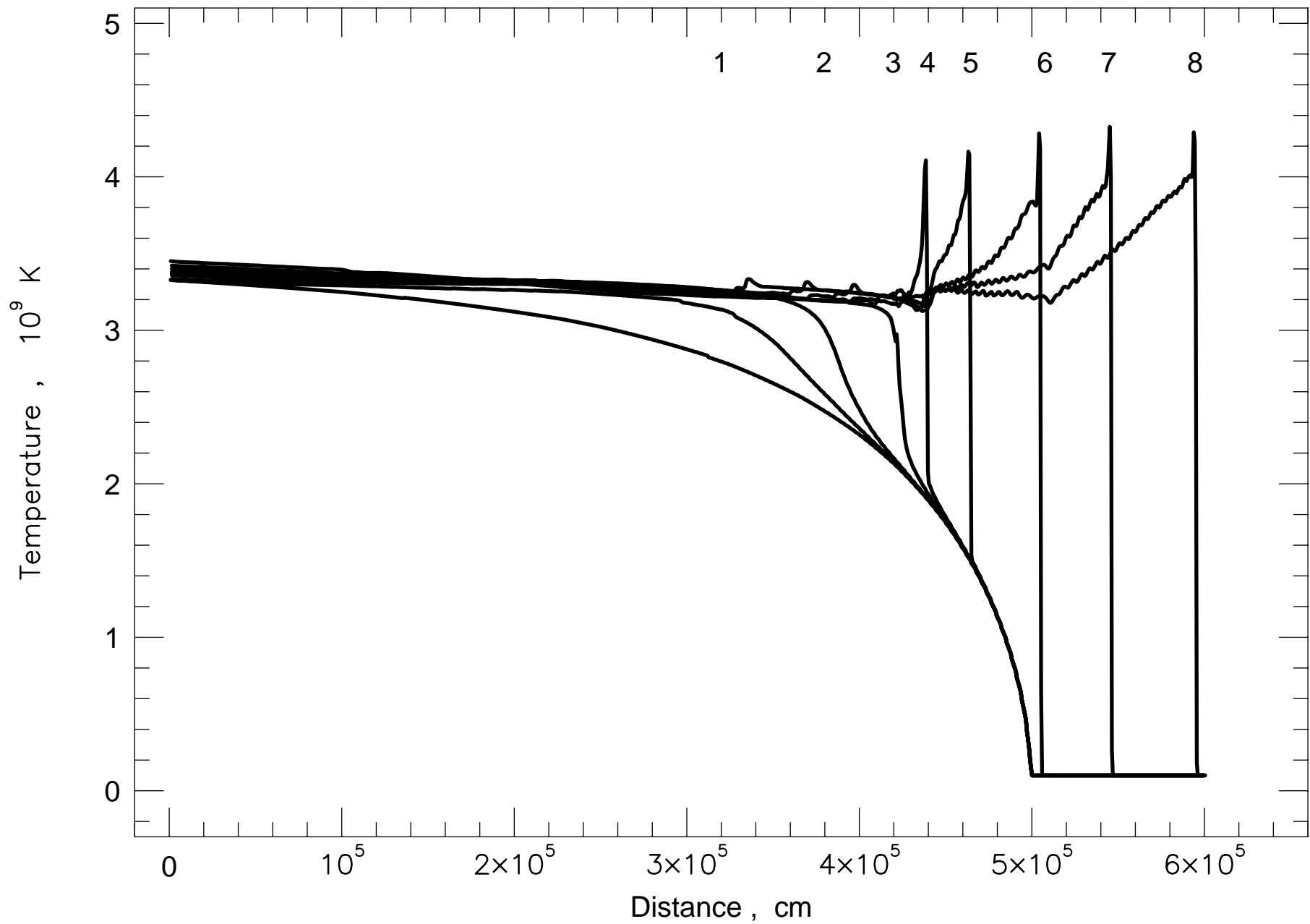
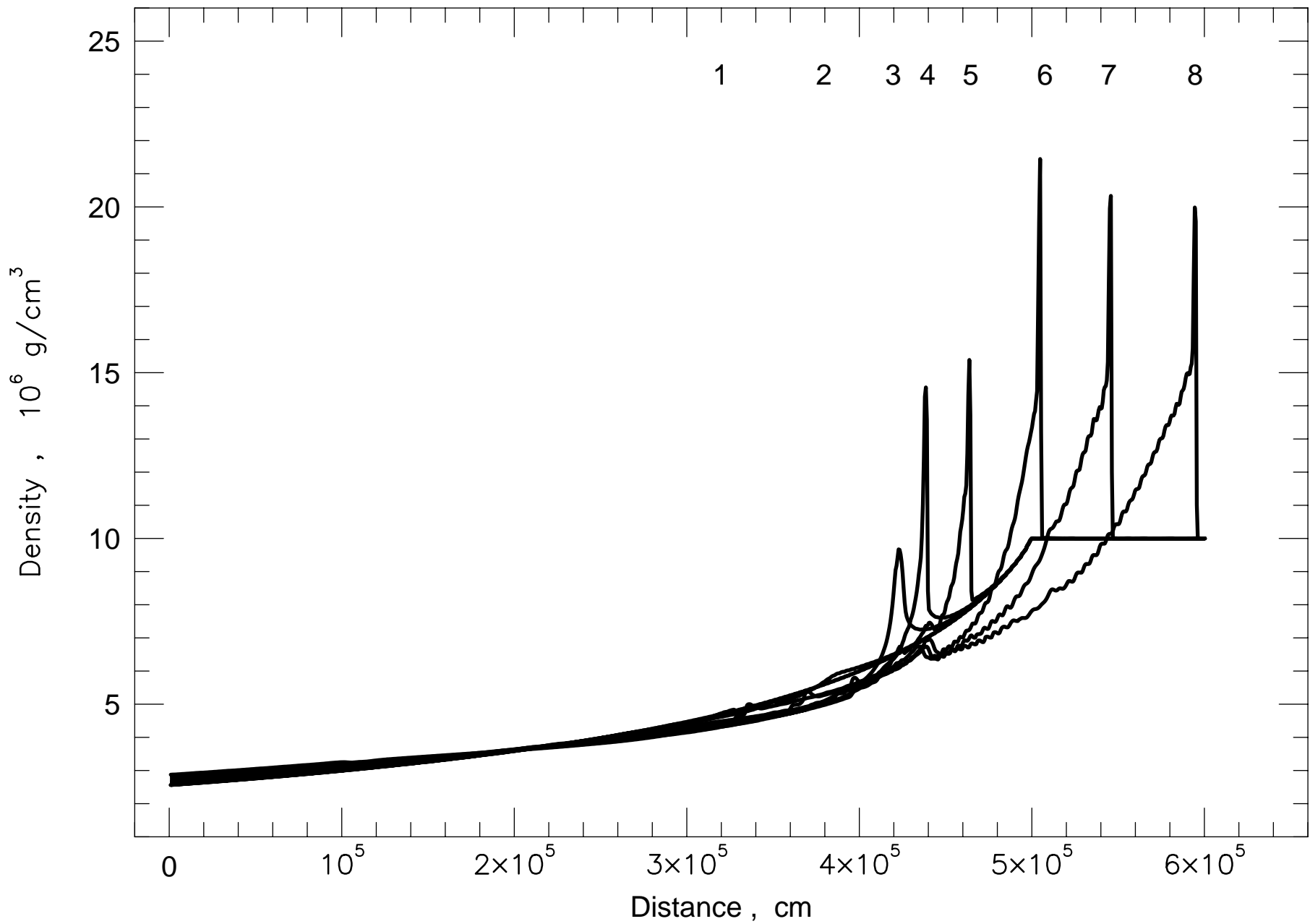
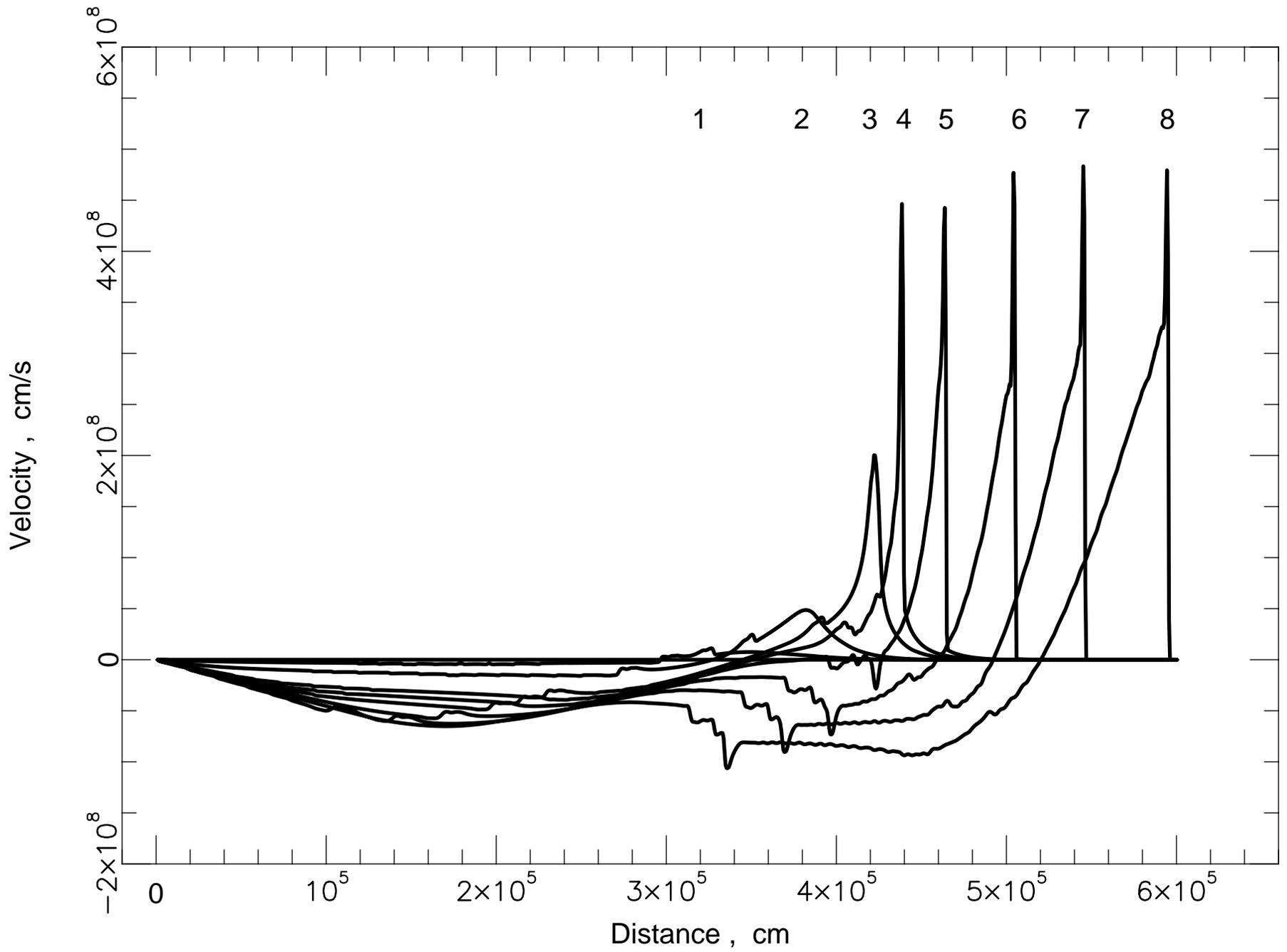
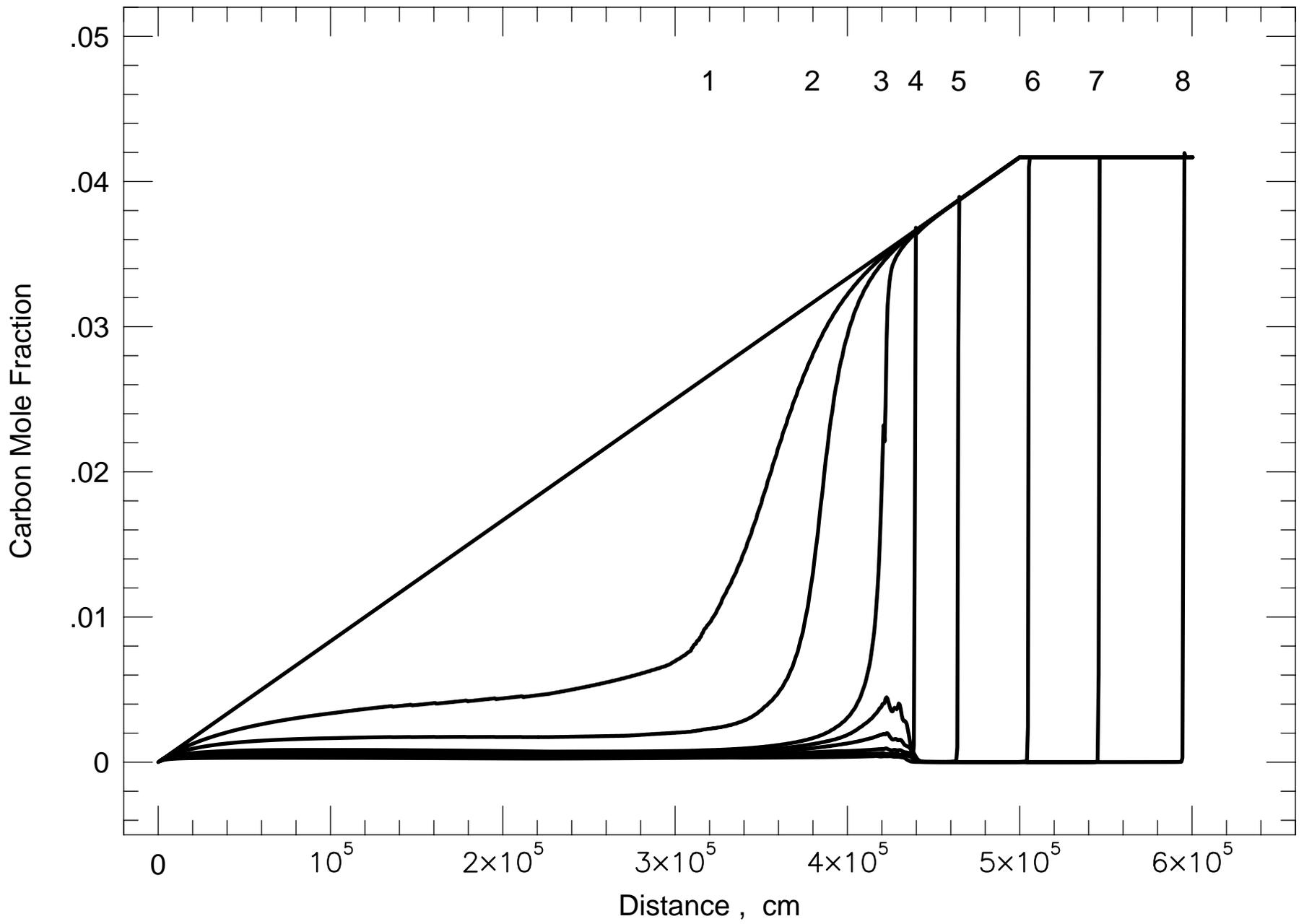
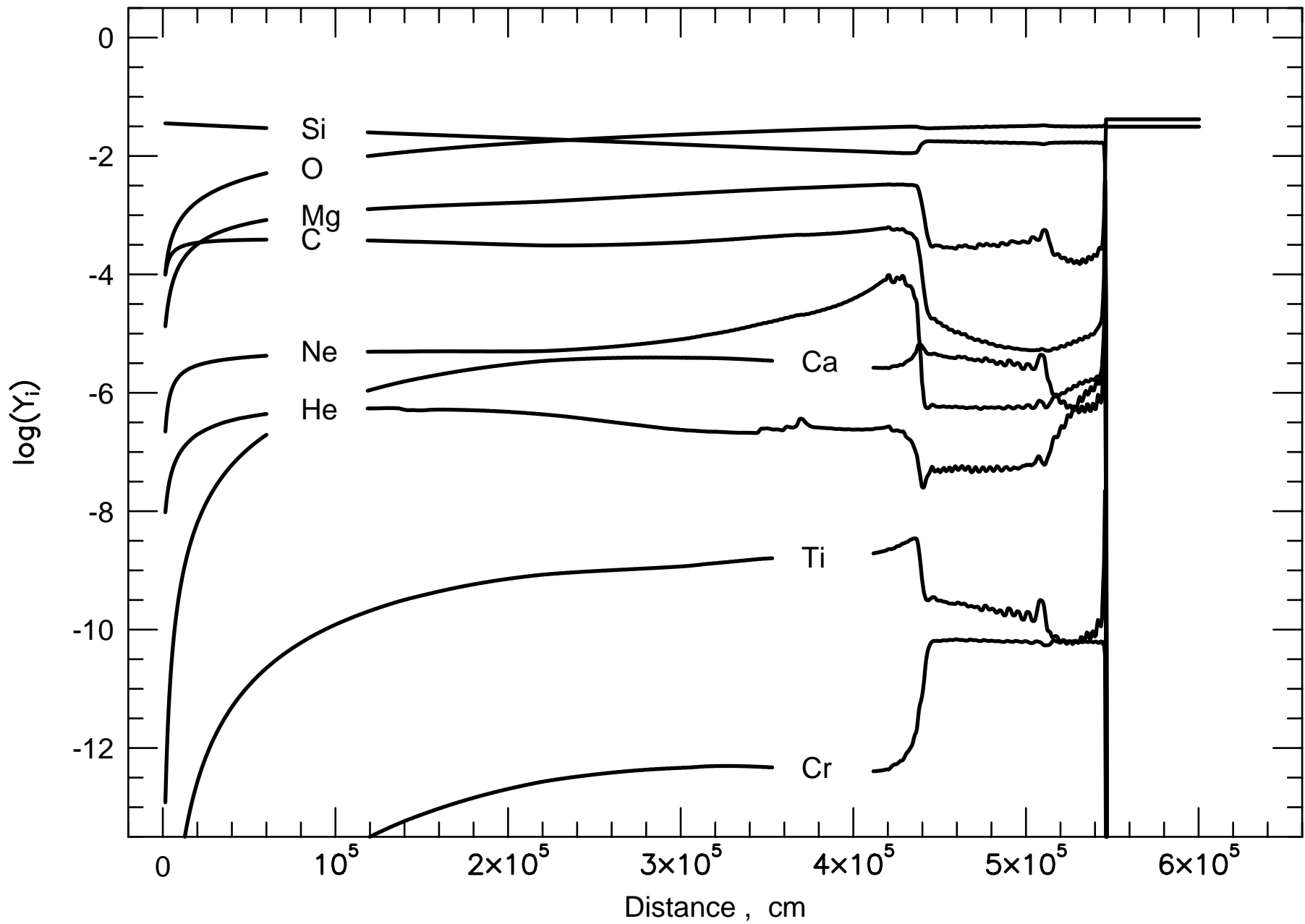
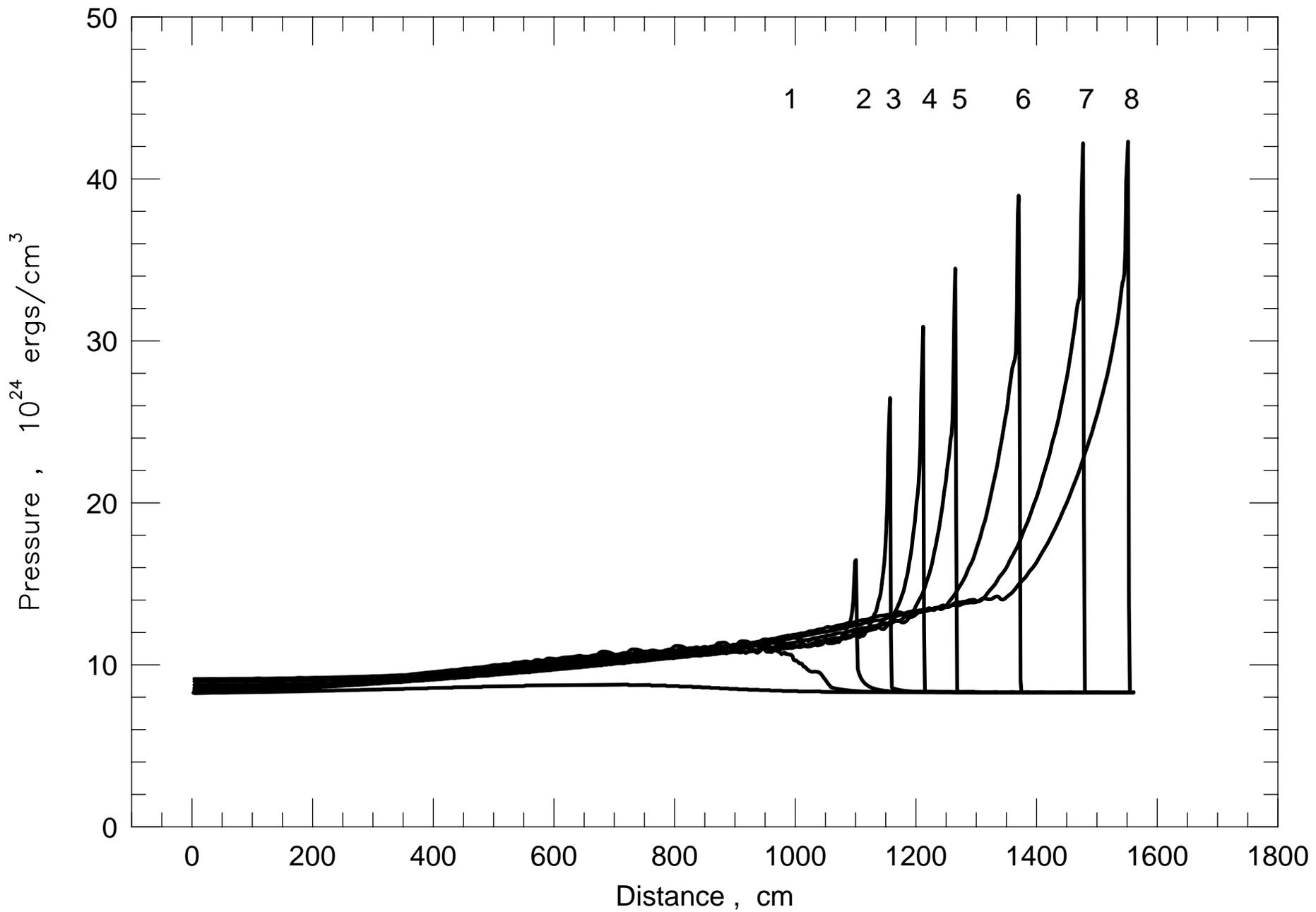
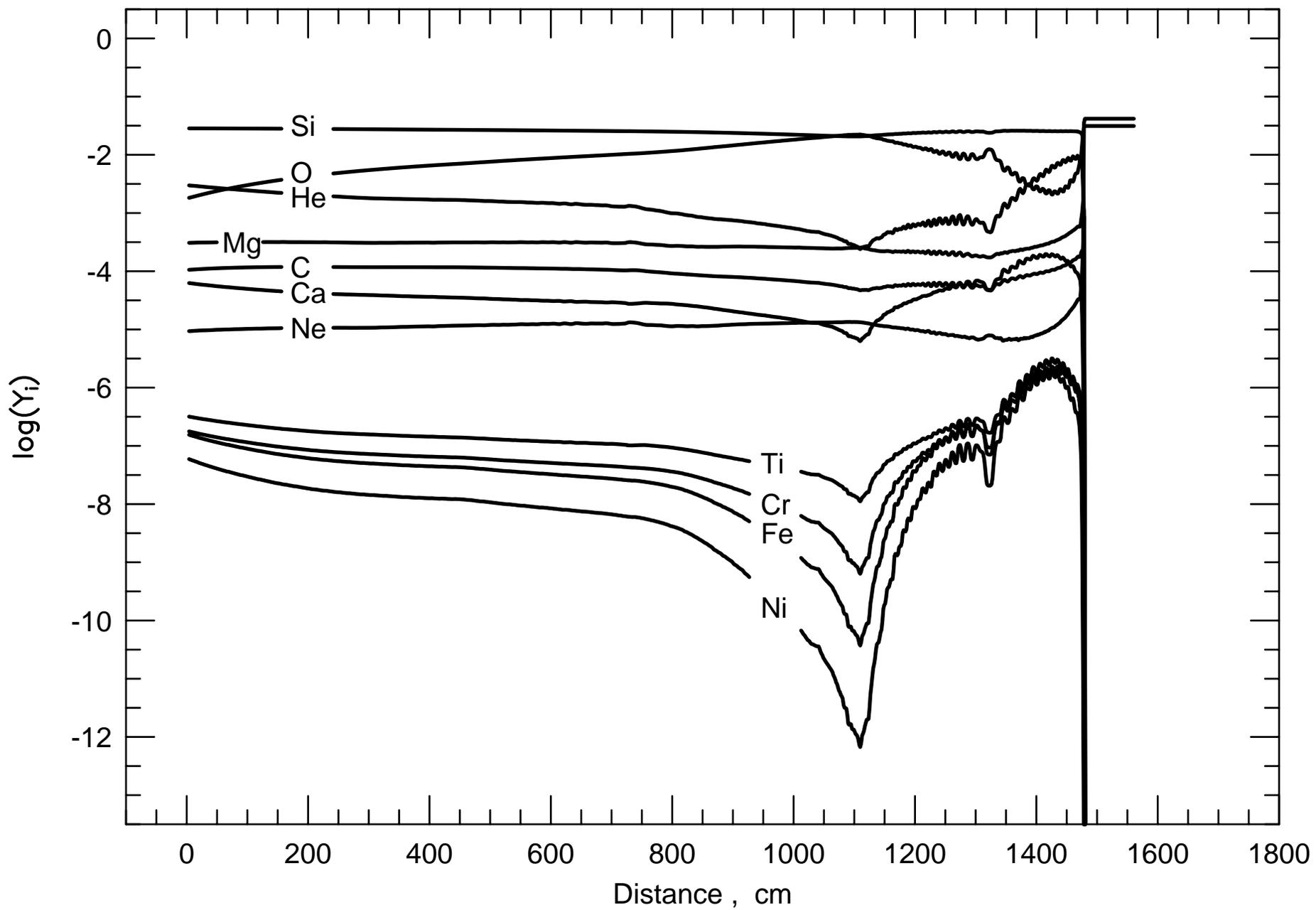
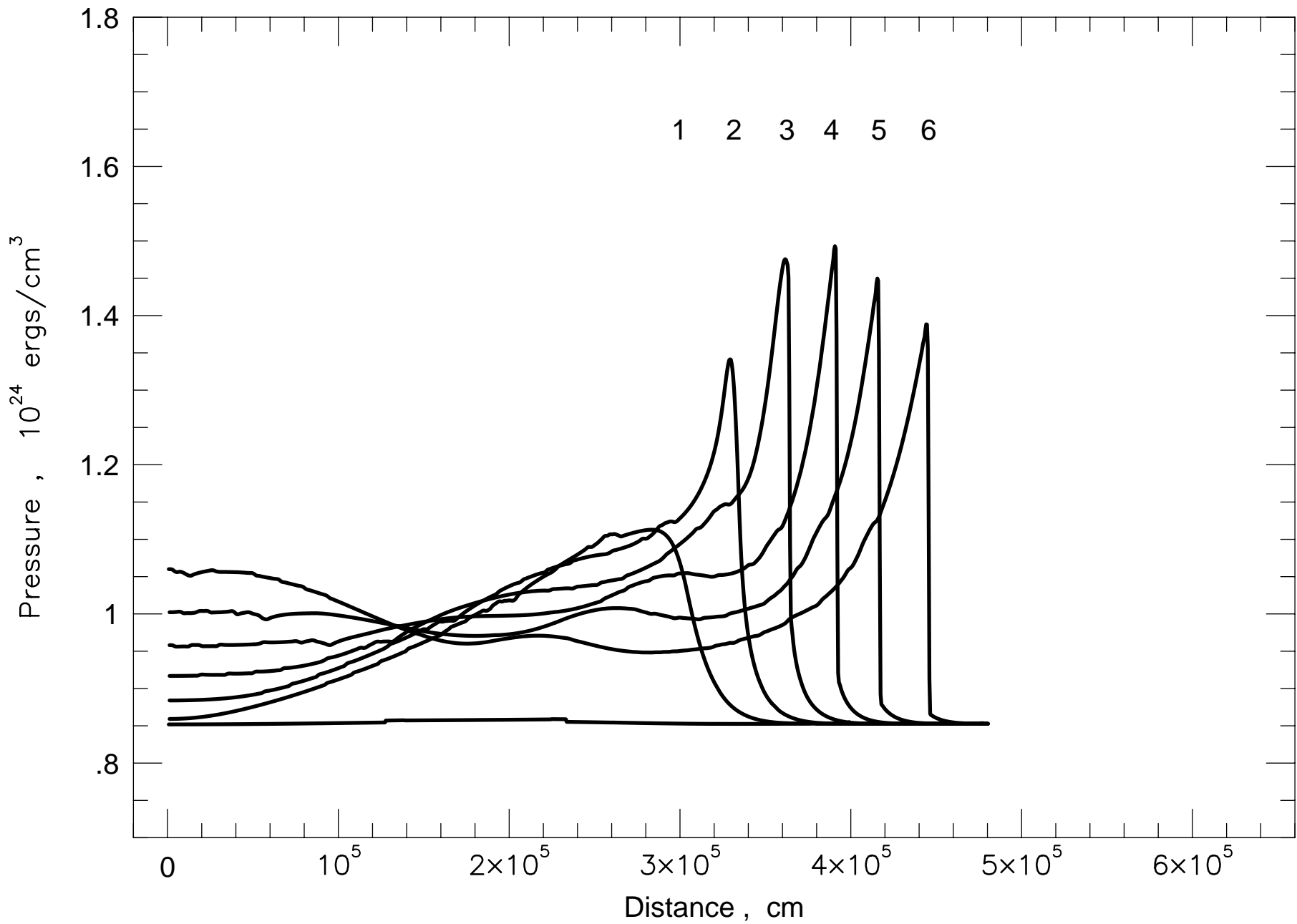
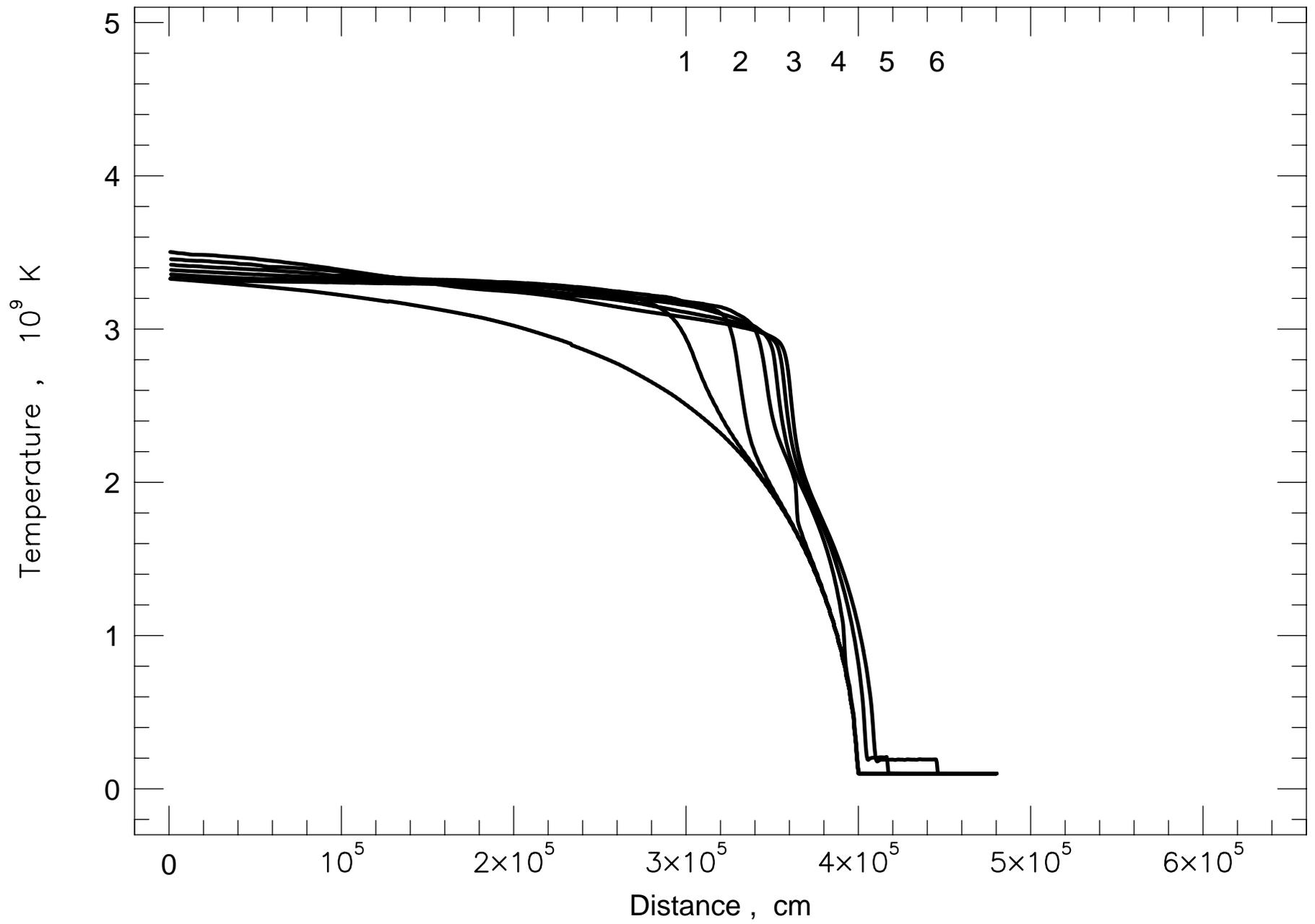
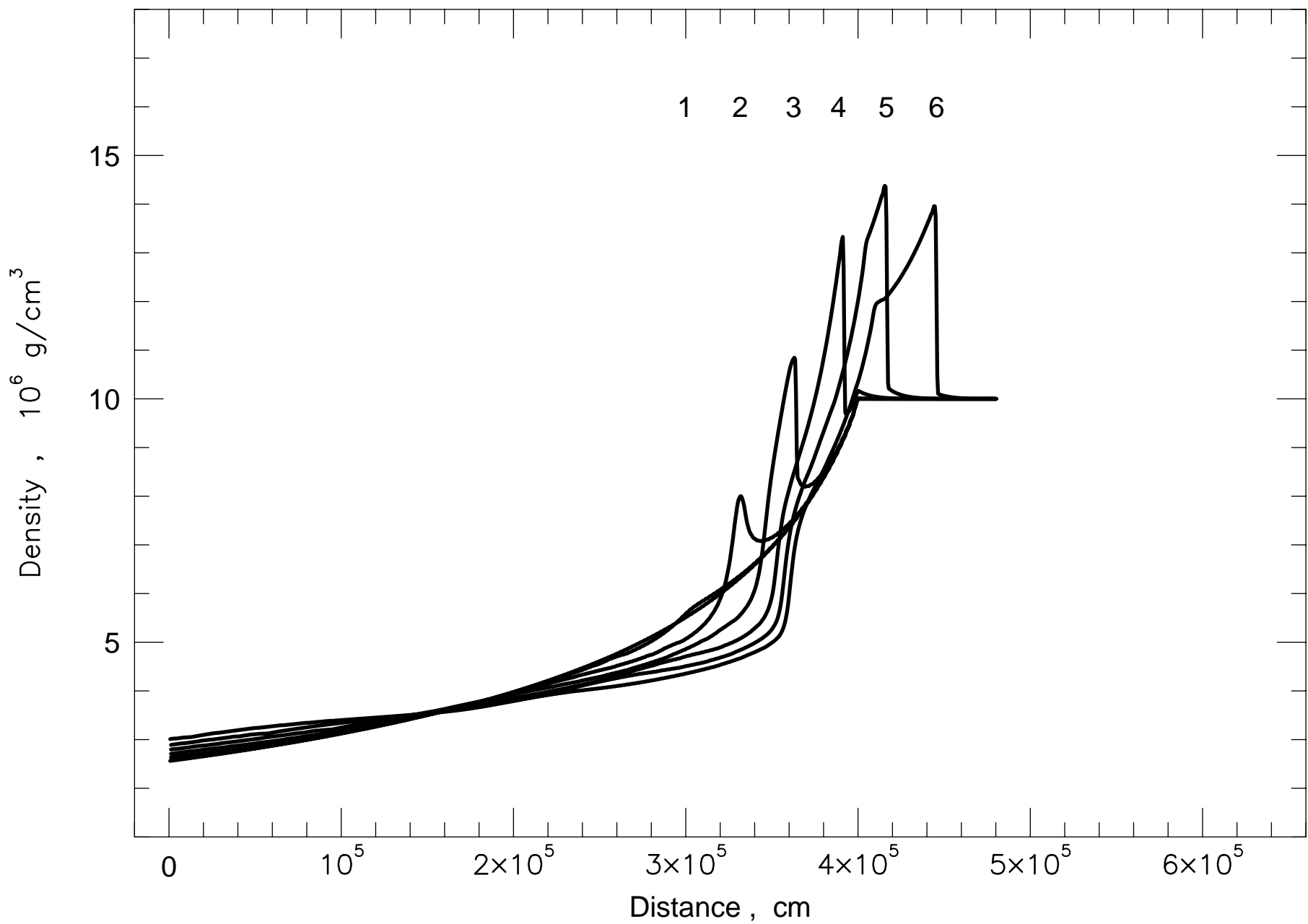
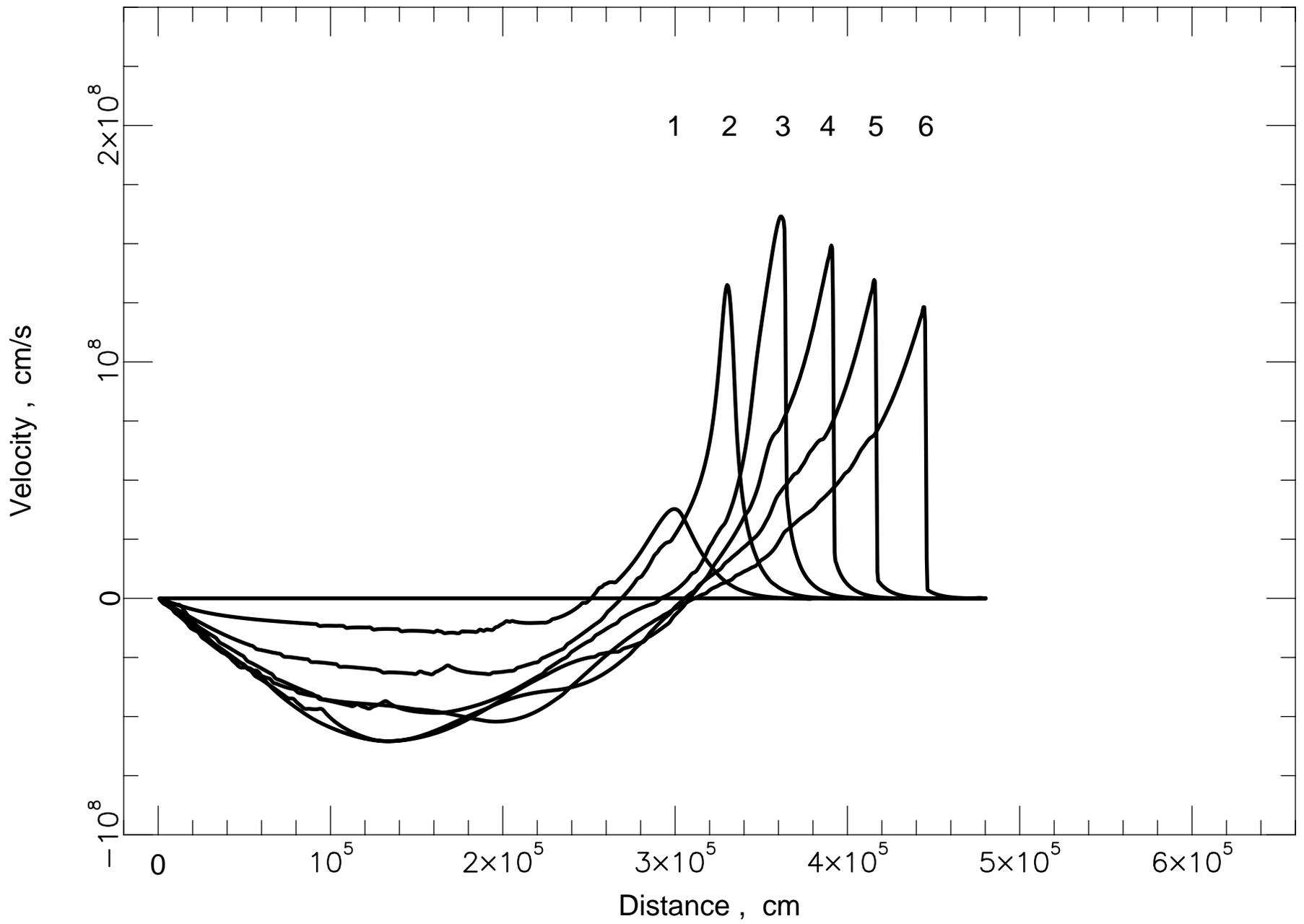
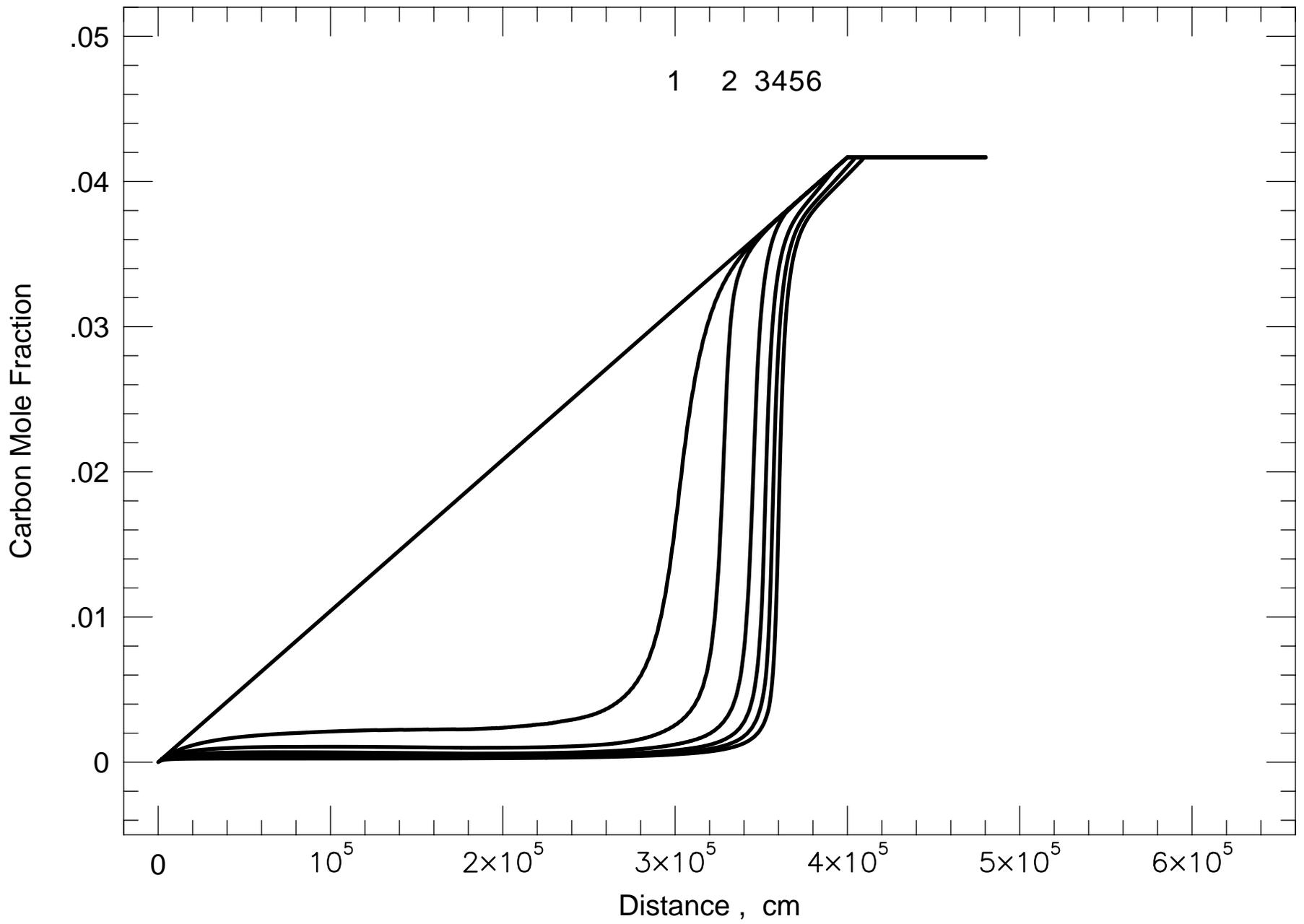
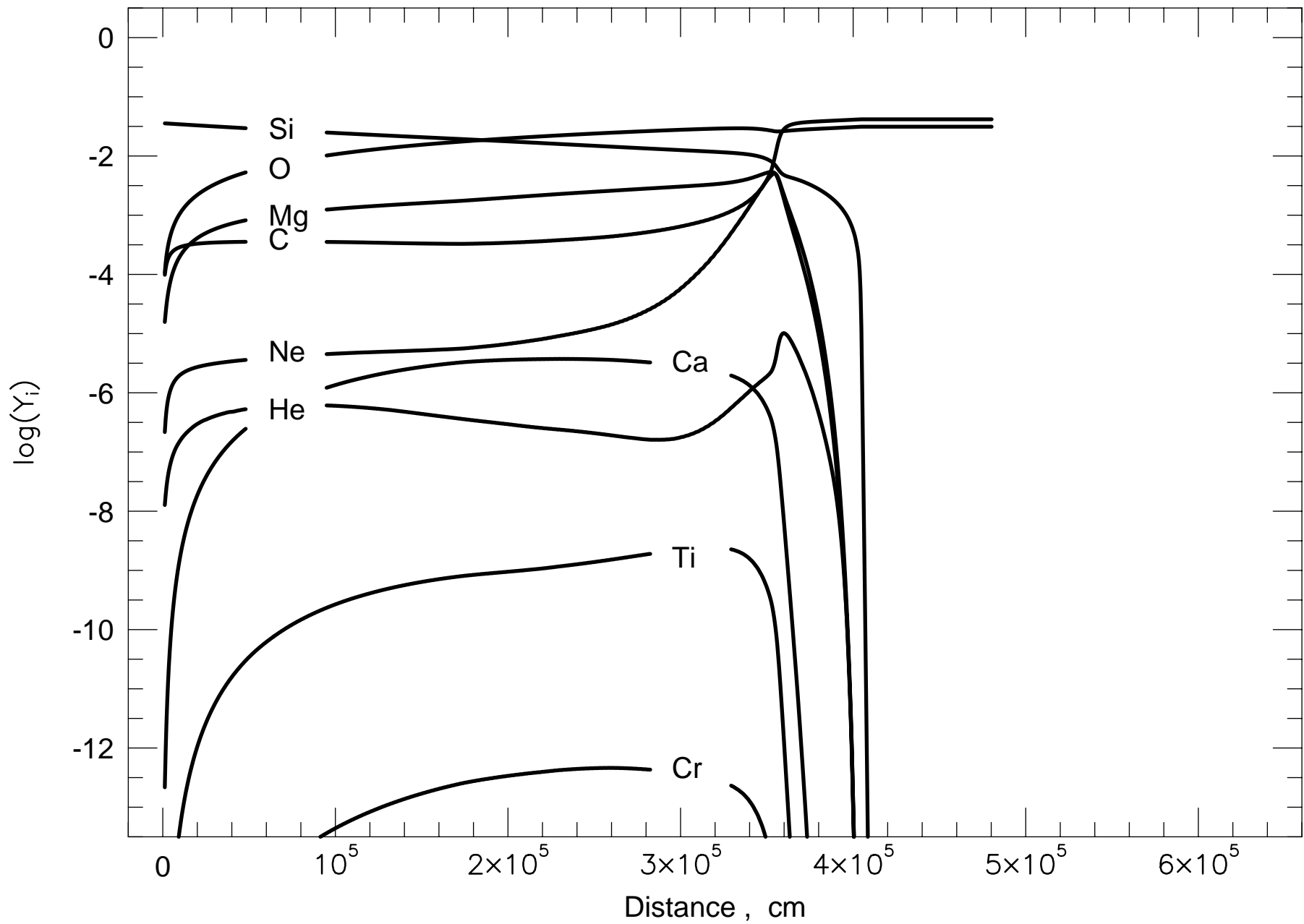
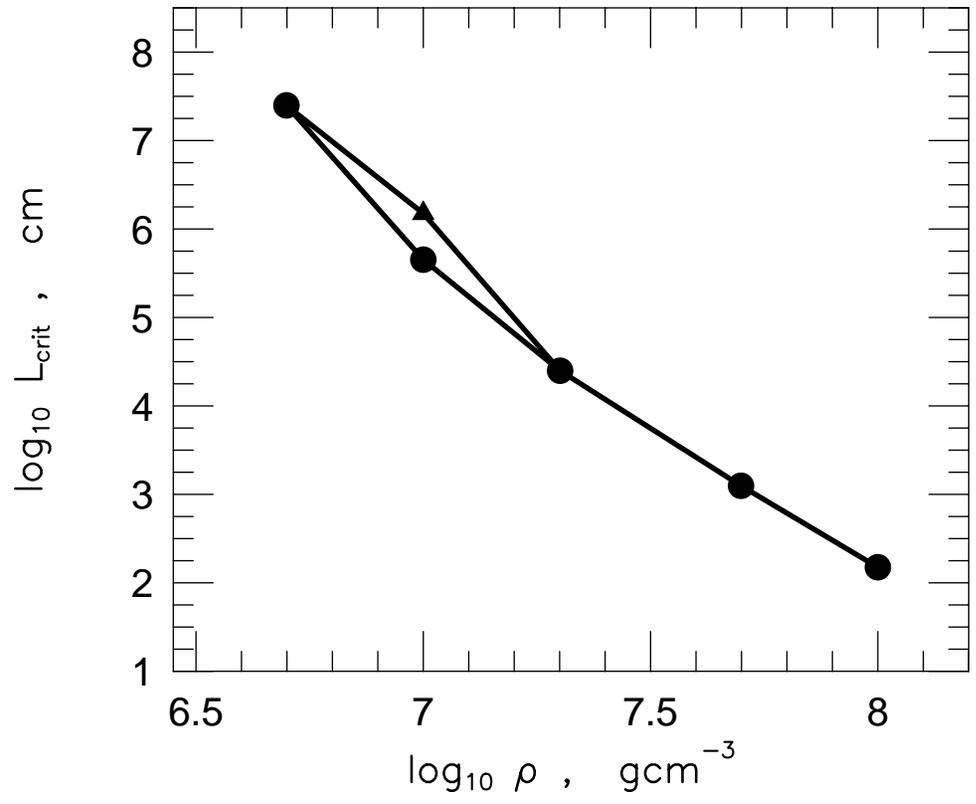
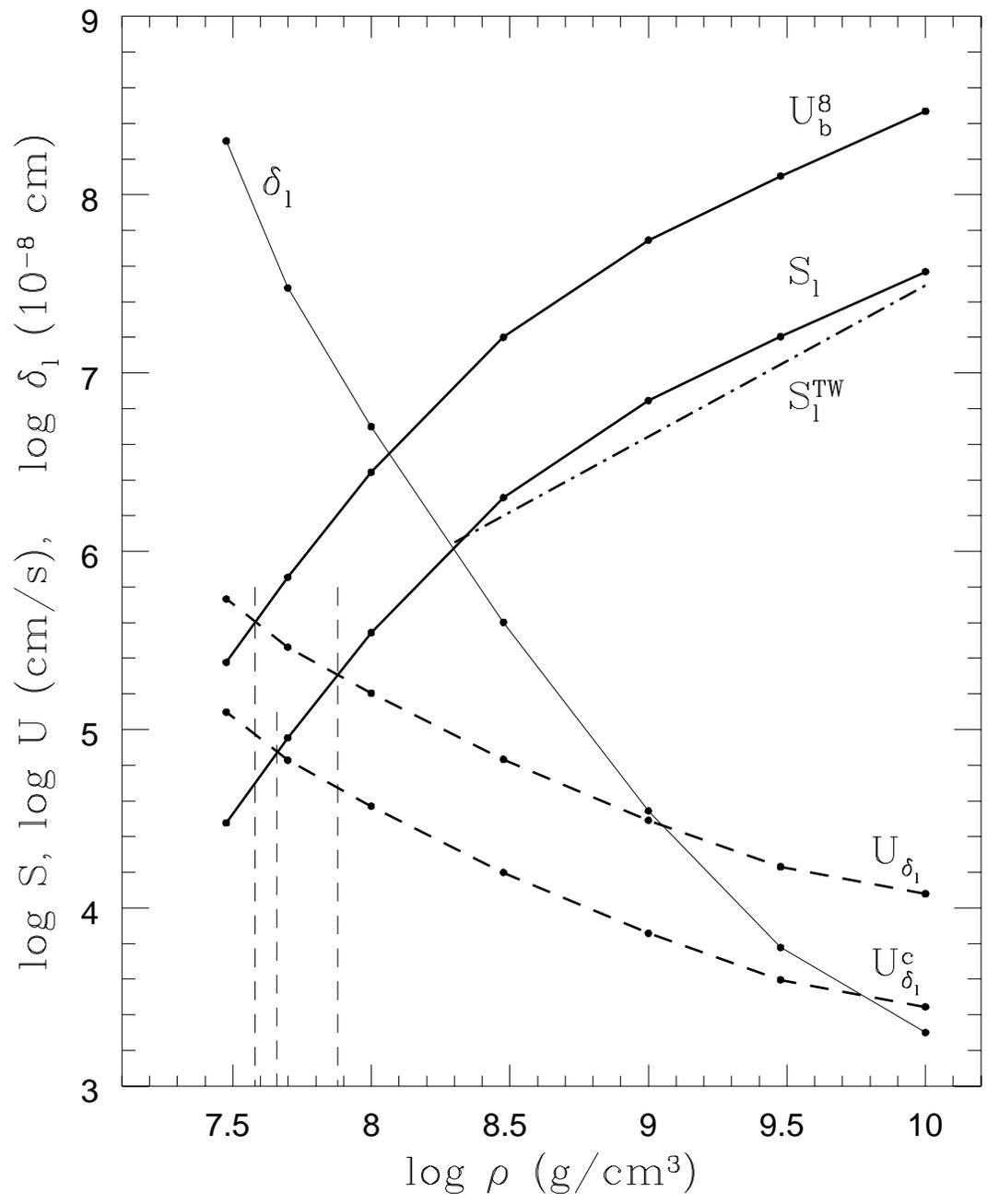
\end{document}